\documentclass[a4paper,fleqn,usenatbib]{mnras}

\usepackage{newtxtext,newtxmath}

\usepackage[T1]{fontenc}
\usepackage{ae,aecompl}

\usepackage{graphicx}	
\usepackage{amsmath}	
\usepackage{amssymb}	
\usepackage{lscape}
\usepackage{rotating}
\usepackage{float}
\hypersetup{colorlinks = true, linkcolor = blue, urlcolor=blue, citecolor=blue} 
\usepackage[normalem]{ulem}
\usepackage{tabularx}
\usepackage{rotating, booktabs}
\usepackage{inputenc}
\usepackage{geometry}
\usepackage{changepage}
\usepackage{color}

\def\msun{\hbox{M$_\odot$}}

\def\vi{$m_{\rm F555W}-m_{\rm F814W} \,$}
\def\i{$m_{\rm F814W} \, $}

\def\v{$m_{\rm F555W} \, $}

\def\uvi{$m_{\rm F275W}-m_{\rm F814W} \,$}
\def\i{$m_{\rm F814W} \, $}

\def\c{$C_{F275W,F343N,F438W} \,$}

\title[Na variations in intermediate age clusters]{Leveraging HST with MUSE: II. Na-abundance variations in intermediate age star clusters}

\author[Martocchia et al.]{S. Martocchia$^{1,2}$, S. Kamann$^{1}$, S. Saracino$^{1}$, C. Usher$^{1,3}$, N. Bastian$^{1}$, M. Rejkuba$^{2}$, \newauthor M. Latour$^{4}$, C. Lardo$^{5}$, I. Cabrera-Ziri$^{6}$\thanks{Hubble Fellow}, S. Dreizler$^{4}$, N. Kacharov$^{7}$, \newauthor V. Kozhurina-Platais$^{8}$, S. Larsen$^{9}$,  S. Mancino$^{2,10,11}$, I. Platais$^{10}$, M. Salaris$^{1}$ \\
$^{1}$Astrophysics Research Institute, Liverpool John Moores University, 146 Brownlow Hill, Liverpool L3 5RF, UK\\
$^{2}$European Southern Observatory, Karl-Schwarzschild-Stra\ss e 2, D-85748 Garching bei M\"unchen, Germany\\
$^{3}$Department of Astronomy, Oskar Klein Centre, Stockholm University, AlbaNova University Centre, SE-106 91 Stockholm, Sweden \\
$^{4}$ Institute for Astrophysics, Georg-August-University G\"ottingen, Friedrich-Hund-Platz 1, D-37077 G\"ottingen, Germany\\
$^{5}$Laboratoire d'astrophysique, \' Ecole Polytechnique F\' ed\' erale de Lausanne (EPFL), Observatoire, 1290, Versoix, Switzerland\\
$^{6}$Harvard-Smithsonian Center for Astrophysics, 60 Garden Street, Cambridge, MA 02138, USA\\
$^{7}$Max-Planck-Institut f\"ur Astronomie, K\"onigstuhl 17, D-69117 Heidelberg, Germany\\
$^{8}$Space Telescope Science Institute, 3700 San Martin Drive, Baltimore, MD 21218, USA\\
$^{9}$Department of Astrophysics/IMAPP, Radboud University, P.O. Box 9010, 6500 GL Nijmegen, The Netherlands\\
$^{10}$Technical University of Munich Cluster of Excellence Universe Boltzmannstr. 2 D-85748 Garching\\
$^{11}$Ludwig-Maximilians-Universit\"at M\"unchen, Geschwister-Scholl-Platz 1, 80539 M\"unchen\\
$^{12}$Department of Physics and Astronomy, Johns Hopkins University, 3400 North Charles Street, Baltimore, MD 21218, USA\\
}
\date{Accepted XXX. Received YYY; in original form ZZZ}

\pubyear{2020}

\begin{document}
\label{firstpage}
\pagerange{\pageref{firstpage}--\pageref{lastpage}}
\maketitle

\begin{abstract}
Ancient ($>$10 Gyr) globular clusters (GCs) show chemical abundance variations in the form of 
patterns among certain elements, e.g. N correlates with Na and anti-correlates with O. Recently, N abundance spreads have also been observed in massive star clusters that are significantly younger than old GCs, down to an age of $\sim$2 Gyr. 
However, 
so far N has been the only element found to vary in such young objects. 
We report here the presence of Na abundance variations in the intermediate age massive star clusters NGC~416 ($\sim$6.5~Gyr old) and Lindsay~1 ($\sim$7.5~Gyr old) in the Small Magellanic Cloud, by combining HST and ESO-VLT MUSE observations.  
Using HST photometry we were able to construct ``chromosome maps'' and separate sub-populations with different N content, in the red giant branch of each cluster. 
MUSE spectra of individual stars belonging to each population were combined, resulting in high signal-to-noise spectra representative of each population, which were compared to search for mean differences in Na. We find a mean abundance variation  of $\Delta$[Na/Fe]$=0.18\pm0.04$ dex for NGC 416 and $\Delta$[Na/Fe]$=0.24\pm0.05$ dex for Lindsay~1. In both clusters we find that the population that is enhanced in N is also enhanced in Na, which is the same pattern to the one observed in ancient GCs.
Furthermore, we detect a bimodal distribution of core-helium burning Red Clump (RC) giants in the UV colour magnitude diagram of NGC 416. A comparison of the stacked MUSE spectra of the two RCs shows the same mean Na abundance difference between the two populations. 
The results reported in this work are a crucial hint that star clusters of a large age range share the same origin: they are the same types of objects, but only separated in age. 
\end{abstract}




\section{Introduction}
\label{sec:intro}

Stellar populations of globular clusters (GCs) are found to host chemical abundance variations in the form of (anti-)correlations among certain light elements, namely He, C, N, O, Na, Al, and sometimes Mg \citep[e.g.,][]{gratton12}. This multiple populations (MPs) phenomenon is observed in every old ($>$10 Gyr) and massive (a few times $10^4$\msun) cluster studied in detail so far. It is only found in small fractions of field stars, consistent with those field stars being originally formed within GCs and subsequently stripped to contribute to galaxy field populations \citep[e.g.,][]{martell11}. 
Over the past $\sim$2 decades many observational and theoretical studies have advanced our understanding of MPs in GCs, however a fully consistent interpretation for the origin and evolution of GCs that includes formation of MPs is still missing (for a recent review see \citealt{bastianlardo18} and references therein).

Early scenarios that were proposed for the origin of the chemical anomalies in clusters were mainly the so-called \textit{self-enrichment} models. As the observed light-element anti-correlations resemble products of high temperature hydrogen burning inside the stars, such models predict that a second generation of stars (P2, enriched in N and Na but depleted in C and O) forms from the material processed by a first generation of stars (P1), together with some amount of unprocessed material which is re-accreted from the surroundings of the cluster (e.g. \citealt{decressin07,dercole08,demink09}). 

However, these scenarios have a number of significant drawbacks that call into question their feasibility, 
the most important ones being: (i) the mass budget problem \citep[e.g.,][]{prantzos06,larsen12}, (ii) the positive correlation between MPs and cluster mass (abundance spread and fraction of P2 stars - \citealt{carretta10,schiavon13,bastian_lardo15,milone17}), (iii) the inability to reproduce some of the detailed observed chemical patterns \citep[e.g.,][]{bastian15,lardo18,cabreraziri19}, (iv) the lack of observational evidence for multiple episodes of star-formation within young massive clusters \citep[e.g.,][]{cabreraziri14,martocchia18b,saracino20} and (v) the observed trends of MP properties with cluster age \citep{martocchia19,li_degrijs19}.

This inability of the early models to explain the observations of MPs has led to the creation of a second generation of models.  \citet{bastian13} proposed that protoplanetary discs around young low-mass stars could sweep up processed material from interacting binary stars (and other high-mass stars) which would then accrete onto the star, causing it to transform from a P1 star to a P2 star.  \citet{gieles18} looked at the formation of super-massive stars in the centres of forming massive clusters which could process material and expel it to be used in further star formation within the clusters.  \citet{breen18} suggested that high energy X-rays originating from the accretion discs around black holes trigger processes leading to the formation of P2 stars. While this generation of models is able to predict some of the observations, still not all can be accounted for, e.g. the discreteness of the variations, the chemical trend of lithium.

A promising line of investigation so far has been to determine whether MPs are found according to certain properties of the cluster.
 Recently, a number of groups have successfully searched for MPs in a variety of galactic environments, including the Magellanic Clouds (MCs, \citealt{mucciarelli09, dalessandro16, niederhofer17a,gilligan19}), the Fornax Dwarf Galaxy \citep{larsen12,larsen14}, M31 \citep{schiavon13,colucci14,sakari16}, the Sagittarius dwarf galaxy (e.g. M54, \citealt{carretta10}) as well as clusters located in local dwarfs and massive galaxies outside the Local Group through integrated spectroscopy \citep[e.g.,][]{larsen_et_al_14,bastian19}. 

Along these lines, we have been carrying out a Hubble Space Telescope (HST) photometric survey along with a European Southern Observatory Very Large Telescope (ESO-VLT) spectroscopic survey to search for chemical anomalies in clusters that are much younger than the typical GC in the MW, i.e. expanding the parameter space of cluster age in the search for MPs. We focused on the MCs because they host many clusters that are as massive as the ancient GCs ($>$ a few times $10^4$\msun), but are much younger, from $\sim$8~Gyr down to a few Myr. 

Within our surveys, we found that many intermediate age clusters (from $\sim2-8$~Gyr) host MPs in the form of N variations in their red giant branch (RGB) stars \citep{niederhofer17a,niederhofer17b,hollyhead17,hollyhead18,hollyhead19,martocchia18a}. Additionally, we observed that older clusters show larger N spreads compared to the younger ones \citep{martocchia19}. This age trend is an extremely important constraint for any model aimed at explaining the formation of MPs. 
However, for a direct comparison with the MPs found in ancient GCs, variations in elements other than N (and, to a certain extent, He, \citealt{chantereau19,lagioia19}) need to be studied.

Previous searches for Na and O variations in young and intermediate age star clusters in the LMC were carried out by \citet{mucciarelli08, mucciarelli14}, and no such variations were found, although the numbers of spectroscopic targets were small. While most of the clusters studied by \citet{mucciarelli08, mucciarelli14} have ages $\lesssim $2~Gyr, which appears to be the minimum age for the presence of N spreads \citep[e.g.,][]{martocchia17,zhang18}, they also did not find a Na-O anti-correlation in the $\sim$2 Gyr cluster NGC~1978, which however shows N spreads \citep{martocchia18a}.

Additionally, studies have been based on integrated light techniques on clusters with a wide range of ages.
\cite{cabreraziri16} and \cite{lardo17} did not find Na and Al anomalies in very young clusters ($<$40 Myr).
Recently, \cite{bastian19} found an extremely enhanced mean Na abundance in G114, a very massive ($>10^7$ \msun) intermediate age ($\sim$3~Gyr) star cluster in the NGC 1316 galaxy but did not find Na enhancement in NGC 1978. The strong Na enhancement in G114 strongly suggests a spread in [Na/Fe] in this cluster. No enhanced Na abundances that would suggest the presence of Na-enriched stars are also reported for two young star clusters ($\sim$600 Myr old) in the merger remnant NGC 7252, even though they are very massive ($\sim10^7-10^8$ \msun, \citealt{bastian20}).

In order to establish whether young and intermediate age massive clusters host the same type of MPs as the ancient clusters, and hence likely share the same origin, we started a VLT/MUSE spectroscopic survey of star clusters in the Magellanic Clouds for which we have available HST photometry.
Here we report the analysis of two intermediate age clusters, NGC 416 ($\sim$6.5~Gyr) and Lindsay 1 ($\sim$7.5~Gyr) in the Small Magellanic Cloud (SMC).
This is the second paper of the series, after \cite{saracino20b}, which aims at searching for differences in light element abundances among the populations of intermediate age clusters in the Magellanic Clouds. 
 
Within this paper we adopt a similar approach to the one reported in \cite{latour19}. We use the so-called ``chromosome map" (ChM, \citealt{milone17}) to separate the two populations with different N abundances. We then combine together all the MUSE spectra belonging to each population, with the aim to investigate difference in mean elemental abundances between the two populations.
Section \ref{sec:obs} describes the observations and both the photometric and spectroscopic reduction. In Section \ref{sec:an} we report on how the different populations were selected according to photometry to look for chemical anomalies. In Section \ref{sec:res} we present the spectroscopic results and comparison with synthetic models for the RGB. In Section \ref{sec:hbs} we report on the red clump (RC, the counterpart of the horizontal branch in clusters at this young ages) morphology of NGC 416. We discuss our results and present our conclusions in Section \ref{sec:disc}.

\begin{table*}
\caption{Main properties of the archival and proprietary HST images for NGC 416 and Lindsay~1.}
\begin{tabular}{c c c c c l l}
\hline
Cluster & Instrument &    Filter  & Year & Exposure Times & Proposal & PI \\
\hline
NGC~416 & UVIS/WFC3 &   F275W  & 2019 & 1500s,1501s,2$\times$1523s, 2$\times$1525s  & 15630 & N.\,Bastian \\  
           & UVIS/WFC3 &   F343N  & 2016 & 3$\times$1200s       & 14069 & N.\,Bastian \\   
           & UVIS/WFC3 &   F438W  & 2016 &  3$\times$375s           & 14069 & N.\,Bastian \\  
           & WFC/ACS   &   F555W  & 2006 &  4$\times$1200s, 2$\times$70s      & 10396 & J. Gallagher \\
           & WFC/ACS   &   F814W  & 2006 &  4$\times$1036s, 2$\times$40s      & 10396 & J. Gallagher \\
\hline 
Lindsay~1 & UVIS/WFC3 &   F275W  & 2019 & 9000s (6 exp)  & 15630 & N.\,Bastian \\  
         & UVIS/WFC3 &   F343N  & 2016 & 4$\times$1200s       & 14069 & N.\,Bastian \\ 
         & UVIS/WFC3 &   F438W  & 2016 &  1040s (3 exp)           & 14069 & N.\,Bastian \\  
         & WFC/ACS   &   F555W  & 2006 &  4$\times$496s, 2$\times$20s      & 10396 & J. Gallagher \\
         & WFC/ACS   &   F814W  & 2006 &  4$\times$474s, 2$\times$10s      & 10396 & J. Gallagher \\
\hline
\end{tabular}
\label{tab:data}
\end{table*}

\section{Observations and data reduction}\label{sec:obs}
\subsection{Photometry}\label{subsec:phot}
The photometric observations of NGC 416 and Lindsay 1 from the HST are composed of both proprietary and archival data. 
The F343N and F438W filters are from the HST/WFC3/UVIS survey of Magellanic Cloud star clusters (proposal GO-14069, P.I. N. Bastian). These observations have been recently complemented with new data in the UV filter F275W, from the proposal GO-15630 (PI: N. Bastian).
ACS/WFC observations in F555W and F814W filters were added for our analysis (GO-10396, PI. J. Gallagher).
The main properties of the WFC3/UVIS and ACS/WFC observations used in the paper, in terms of exposure times in the different filters, are summarised in Table \ref{tab:data}.

The photometric analysis has been performed with \texttt{DAOPHOT IV} \citep{stetson87} on images processed, flat-fielded, bias subtracted, and corrected for Charge Transfer Efficiency losses by standard HST pipelines ($\_{flc}$ images). As a first step, few hundreds of stars have been selected in each image and chip in order to model the point spread function (PSF), by considering a 10-pixel aperture. The PSF models were chosen on the basis of a $\chi^2$ statistic and, on average, the best-fit has been provided by a Moffat function \citep{moffat69}. These models were finally applied to all the sources detected at more than $3\sigma$ from the background level in each image. Then, we built a master catalog with stars detected in at least half of the available images per filter. In some cases a less restrictive criterion has been adopted in order to cover the gap between the two chips. At the corresponding positions of these stars, the photometric fit was forced in all the other frames by using {\texttt{DAOPHOT IV/ALLFRAME} \citep{stetson94}}. Finally, the magnitude and photometric error of each star has been estimated as the weighted mean and standard deviation of the magnitudes measured in the individual images.

Instrumental magnitudes have been then transformed to the VEGAMAG system by using the zero-point values quoted both on the WFC3 and ACS websites (at the aperture size), as well as appropriate aperture corrections at a radius of 10 pixels.
Instrumental positions were first corrected for geometric distortions \citep{meurer03,bellini11} and have been transformed to the absolute coordinate system (RA, Dec) by using the stars in common with the Gaia Data Release 2 (DR2, \citealp{gaia2016,gaia2018}) and by means of the cross-correlation software \texttt{CataXcorr} \citep{montegriffo1995}. This software has been also used to combine the final WFC3 and ACS catalogues of the clusters.

\subsection{MUSE data}\label{subsec:muse}

\begin{table}
\caption{MUSE observations taken for NGC 416 and Lindsay~1.}
\begin{tabular}{c c c c}
\hline
Cluster & Date & Exposure Times & Image Quality  \\
\hline
Lindsay~1 & 2019-10-31 & $4\times660$~s & $1.1\arcsec$ \\
          & 2019-11-03 & $4\times660$~s & $1.1\arcsec$ \\
          & 2019-11-25 & $4\times660$~s & $0.9\arcsec$ \\
          & 2019-12-01 & $4\times660$~s & $0.8\arcsec$ \\
NGC~416 & 2019-11-03 & $4\times660$~s & $0.8\arcsec$ \\
        & 2019-11-30 & $4\times660$~s & $0.9\arcsec$ \\
\hline
\end{tabular}
\label{tab:muse_data}
\end{table}

MUSE observations of NGC 416 and Lindsay 1 were obtained between October and December 2019 in the course of program 0104.D-0257 (PI: Kamann), using the Wide Field Mode (WFM) of the instrument without adaptive optics support. Each observation consisted of four exposures and derotator offsets of $90^\circ$ were applied in between exposures. The individual observations are listed in Table~\ref{tab:muse_data}.

The data reduction was performed using version 2.8.1 of the standard MUSE pipeline \citep{weilbacher2020}. The basic reduction steps (bias removal, spectrum tracing, flat fielding, and wavelength calibration) were executed on a per-integral field unit (IFU) basis. Afterwards, the data from the 24 IFUs were combined and a sky subtraction was performed. In the final step, all four exposures taken during a single observing block were combined into a final data cube. We measured the effective image quality of our observations on the white-light images created by collapsing the final data cubes along the spectral axis. The values, included in Table~\ref{tab:muse_data}, show that the first two observations of Lindsay~1 were taken under worse seeing conditions, which limited the number of spectra we obtained from the data.

\begin{figure*}
\centering
\includegraphics[scale=0.55]{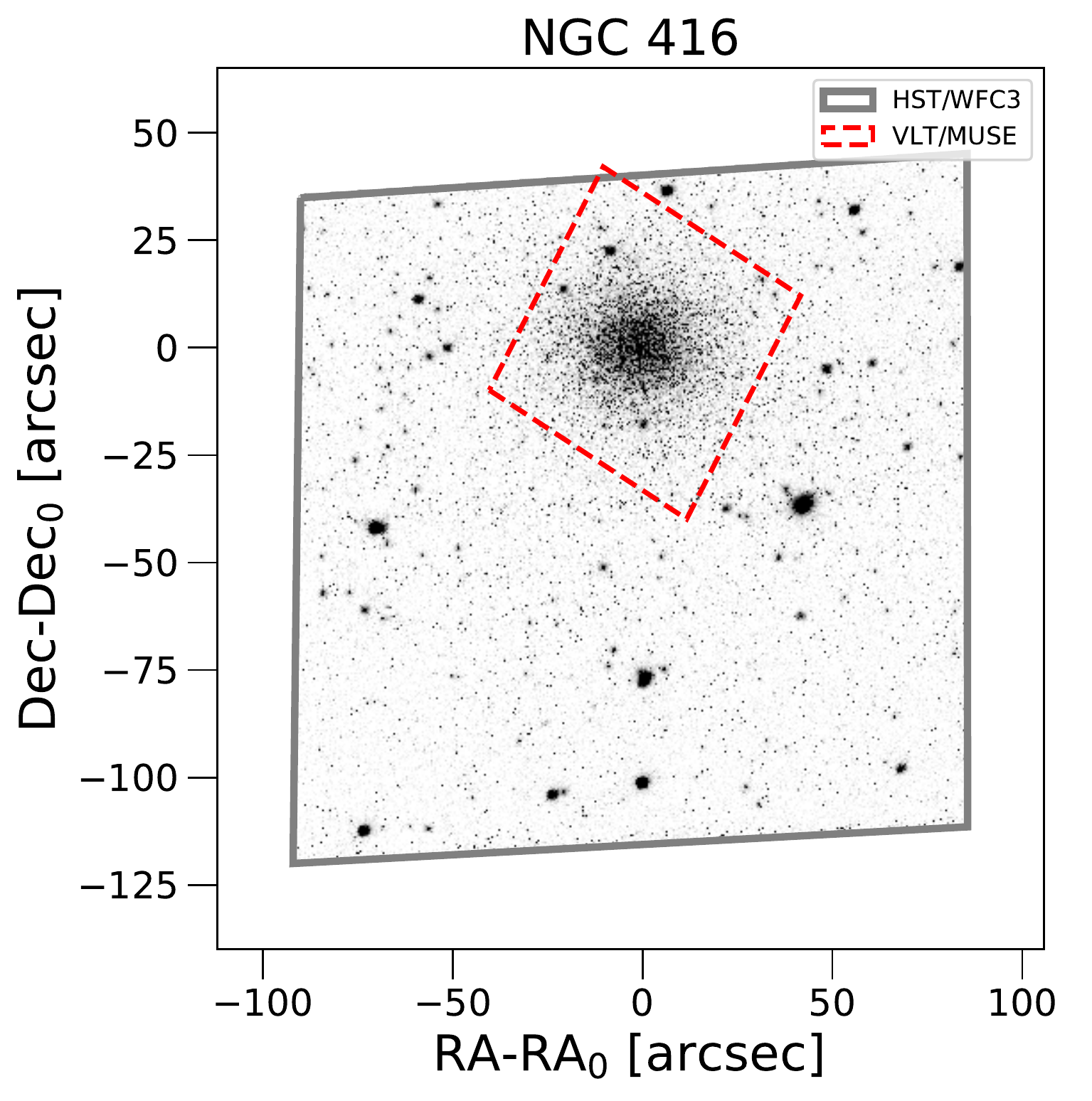}
\includegraphics[scale=0.55]{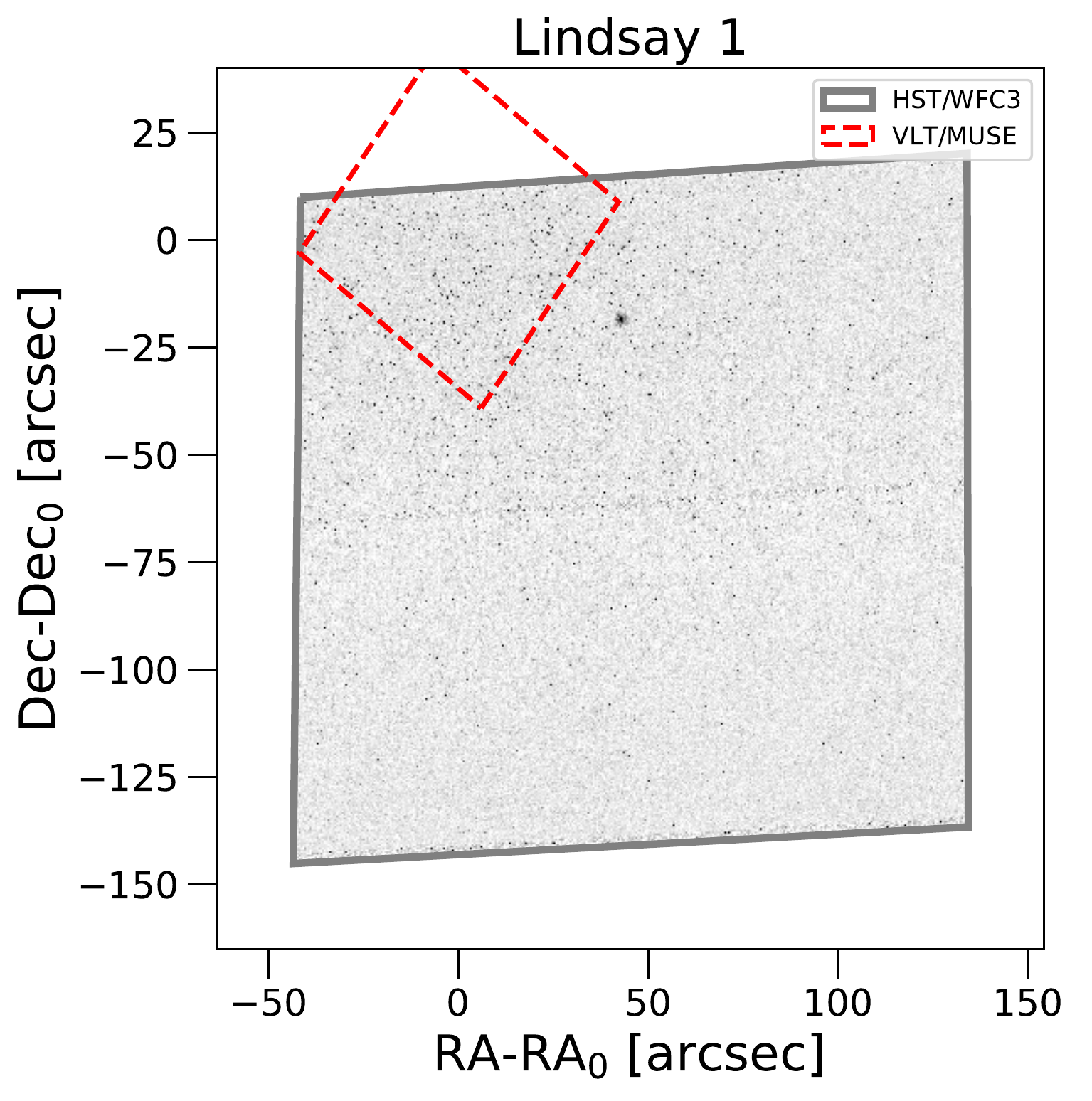}
\caption{HST/WFC3 images of NGC~416 and Lindsay~1 in F336W and F343N filters, respectively. The red dashed box represents the VLT/MUSE field of view.} 
\label{fig:maps}
\end{figure*}

We used \textsc{PampelMuse} \citep{kamann13,kamann18} to extract individual spectra from the MUSE cubes, using the HST photometric catalogue introduced in Sect.~\ref{subsec:phot} as a reference. For NGC~416, we extracted a total of 5064 spectra for 3236 stars and for Lindsay~1, 2424 spectra for 929 stars were extracted. The larger number of spectra extracted from the NGC 416 data can be explained by the higher density of this cluster compared to Lindsay~1 in addition to better seeing conditions during the observations. In this paper we focus on the subset of stars with (a) a spectroscopic Signal-to-Noise Ratio (SNR) $>20$, and (b) a photometrically determined population membership, consisting of $145$ RGB and $149$ RC stars in NGC~416 and $25$ RGB stars in Lindsay~1. Unfortunately, the number of stars used in this analysis for Lindsay~1 is quite small in the end. Besides the reasons reported above, this is also due to an offset between the HST and VLT/MUSE observation field of views. This is shown in Fig. \ref{fig:maps}, where the HST/WFC3 drizzled images of NGC~416 (left panel) and Lindsay~1 (right) are displayed\footnote{The images were downloaded from \url{https://archive.stsci.edu/hst/}}. These are in the F336W filter for NGC~416 and F343N for Lindsay~1. More generally, the limiting filters for Lindsay~1 are the F343N and F438W. The red dashed box indicates the VLT/MUSE field of view.

The reduced spectra were processed with \textsc{Spexxy}\footnote{\url{https://github.com/thusser/spexxy}} \citep{husser16},
which determines radial velocities, metallicities, and effective temperatures via full spectrum fits by using synthetic \textsc{GLib} \citep{husser13} templates (see Table \ref{tab:n416_rgb}, \ref{tab:l1}, \ref{tab:n416_hb}). 
To determine the surface gravities, we compared the HST photometry with the isochrones reported in Section \ref{subsec:models}, because the estimation of $\log g$ from MUSE spectra is challenging \citep{husser16}.

Telluric absorption from the Earth atmosphere in the spectra were corrected by dividing each spectrum by the telluric model obtained in the full-spectrum fit. Additionally, the fit yields a polynomial that represents the
difference between the observed spectrum and the model. We additionally divided each spectrum by its polynomial in order to homogenize the continua.

\begin{figure*}
\centering
\includegraphics[scale=0.6]{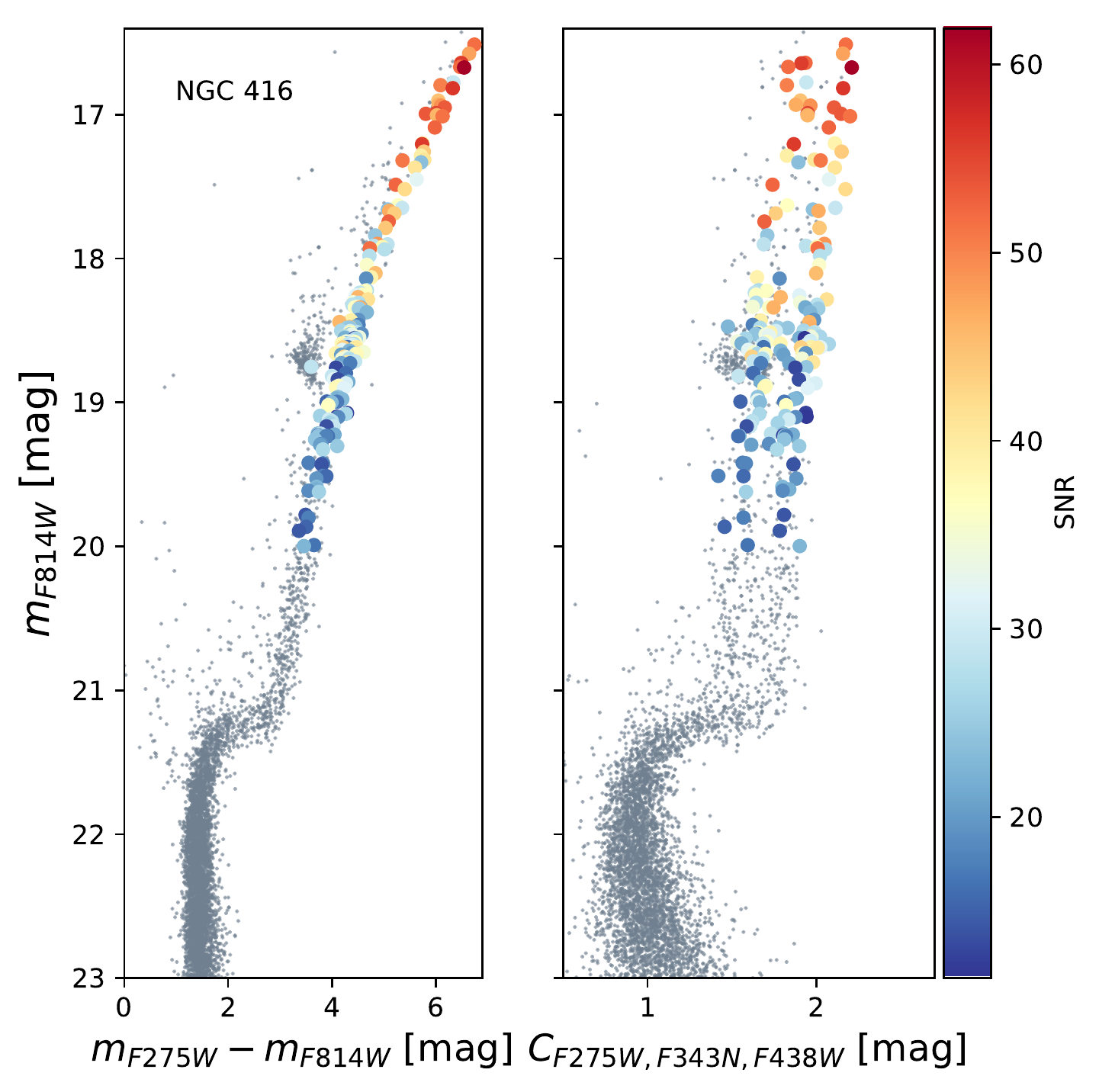}
\includegraphics[scale=0.56]{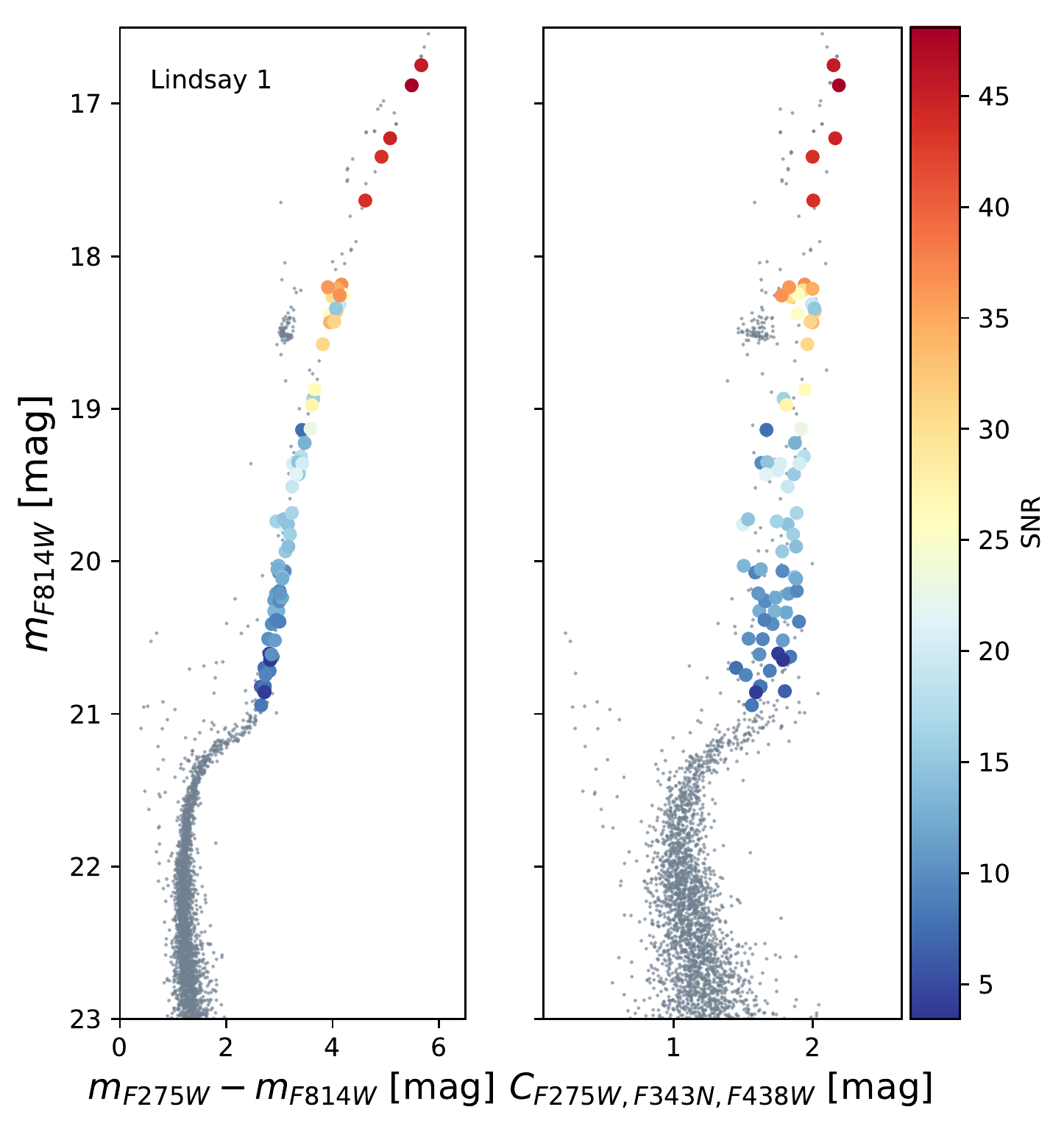}
\caption{\textit{Left panels}:  \uvi vs. \i CMD and \c vs. \i CMD of NGC~416; \textit{Right panels}: \uvi vs. \i CMD and \c vs. \i CMD of Lindsay~1. In all panels, MUSE RGB targets are colour-coded by their SNR.} 
\label{fig:cmds}
\end{figure*}

\section{Analysis}\label{sec:an}

\subsection{Differential reddening}\label{subsec:dr}

We corrected our photometric catalogues for differential reddening (DR) by using the method reported in \cite{milone12}. 

We first identified the reddening vector in the \vi vs. \v colour-magnitude diagrams (CMDs), by adopting the extinction coefficients by \cite{casagrande14}. We then selected main-sequence stars in the magnitude range $22.2\lesssim$\v$\lesssim 23.8$ for NGC 416 and $22.2\lesssim$\v$\lesssim 24.2$ for Lindsay~1 as reference stars. We translated all the cluster stars in a ``rotated parameter space'' where the extinction vector is parallel to the x-axis and we then defined a fiducial line for the previously selected stars. 
At each star in the catalogues, we assigned a distance along the x-axis ($\Delta$D), by computing the mean of the 25 nearest (in space) stars within our selection. We applied a 2$\sigma$-clipping to remove clear outliers from the distribution, where $\sigma$ represents the x-difference between the stars and the fiducial line. It is worth to mention that by changing the number of neighbour stars (from 20 to 30), we obtain very similar results. Finally, the local DR $\delta E(B-V)$ for each individual star is obtained by moving back into the classical \vi vs. \v CMD through the same extinction vectors.

As stated in \cite{saracino19}, for Lindsay~1 the DR is very low, with a mean $\delta E(B-V)\sim$0.003 mag. NGC 416 is affected more strongly by DR (see \citealt{niederhofer17b}). We find a maximum $\delta E(B-V)\sim$0.02 mag and a minimum of $\delta E(B-V)\sim -$0.04 mag, with a mean of $\delta E(B-V)\sim -$0.01 mag. 
Hereafter, we will use the DR corrected photometric catalogues.

\subsection{Chromosome Map}\label{subsec:chm}
RGB stars were selected from the \uvi vs. \i CMD, from \i=21 mag up to the tip of the RGB around \i$\sim$16.5 mag for both clusters. The \uvi vs. \i CMD of NGC 416 is shown in the left panel of Fig. \ref{fig:cmds}. 
The second panel shows the \c$\equiv(F275W-F343N)-(F343N-F814W)$ vs. \i CMD of NGC~416, where a split RGB is clearly visible. 
We also show the \uvi vs. \i and \c vs. \i CMDs of Lindsay~1 in the right panels of Fig. \ref{fig:cmds} (see also \citealt{saracino19}). MUSE spectroscopic RGB targets overplotted with large dots in all panels are colour-coded by their SNR. 

Next, we built the chromosome map (ChM) of both clusters. The ChM was introduced by \cite{milone17} and shown to effectively separate MPs in a given cluster. The ChM represents a (pseudo)colour-colour diagram which is mainly sensitive to He variations on the x$-$axis (\uvi), while it is sensitive to N variations on the y$-$axis (\c - e.g., \citealt[][]{lardo18}). Here we substitute the F343N filter for the F336W filter used by \citet{milone17} due to its increased sensitivity to N-spreads. 
The use of a different filter does not affect our results (see e.g. \citealt{saracino19,saracino20}).  We then used the ChM to separate the two populations present in NGC~416 and Lindsay~1.

\begin{figure*}
\centering
\includegraphics[scale=0.38]{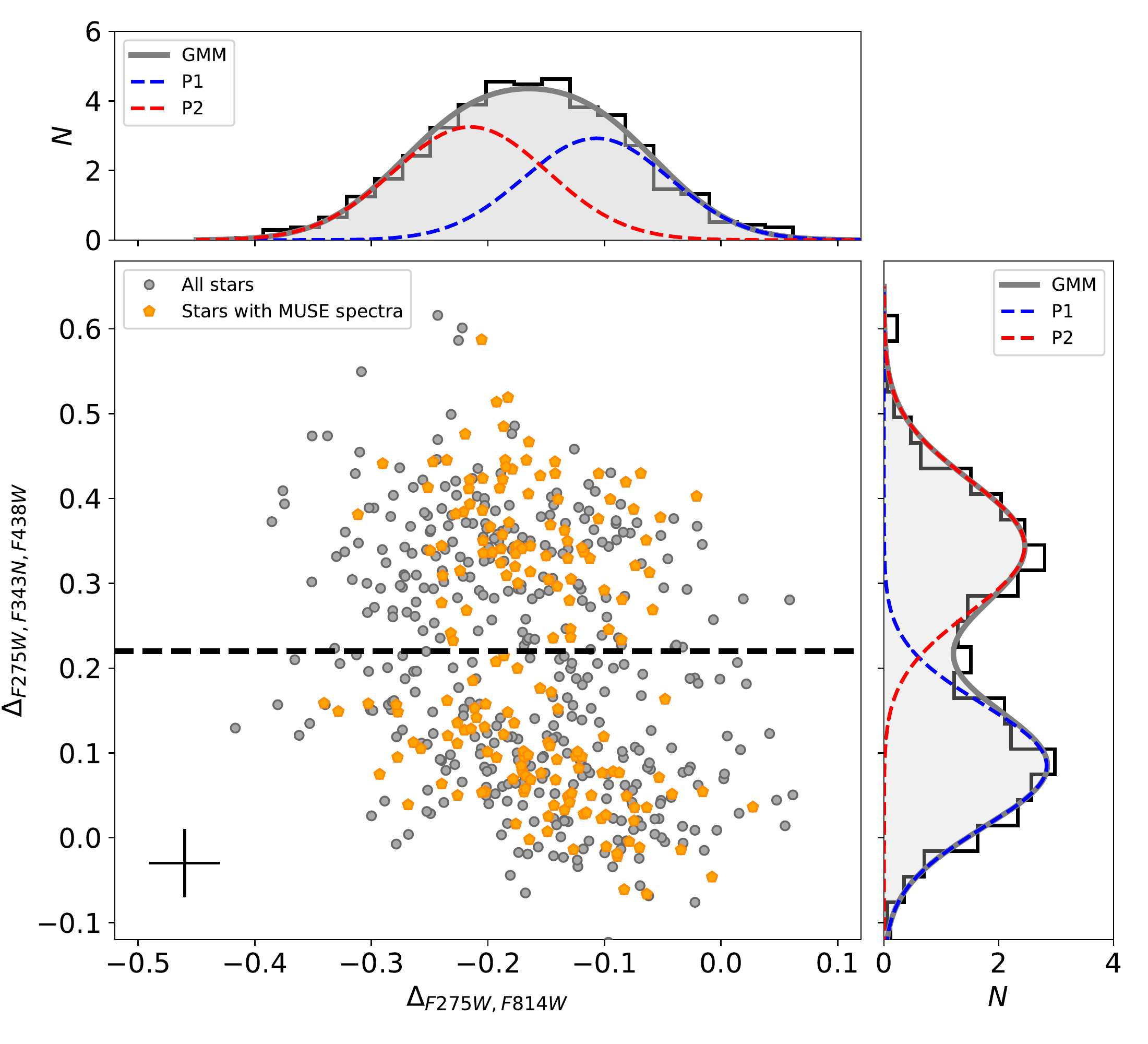}
\includegraphics[scale=0.38]{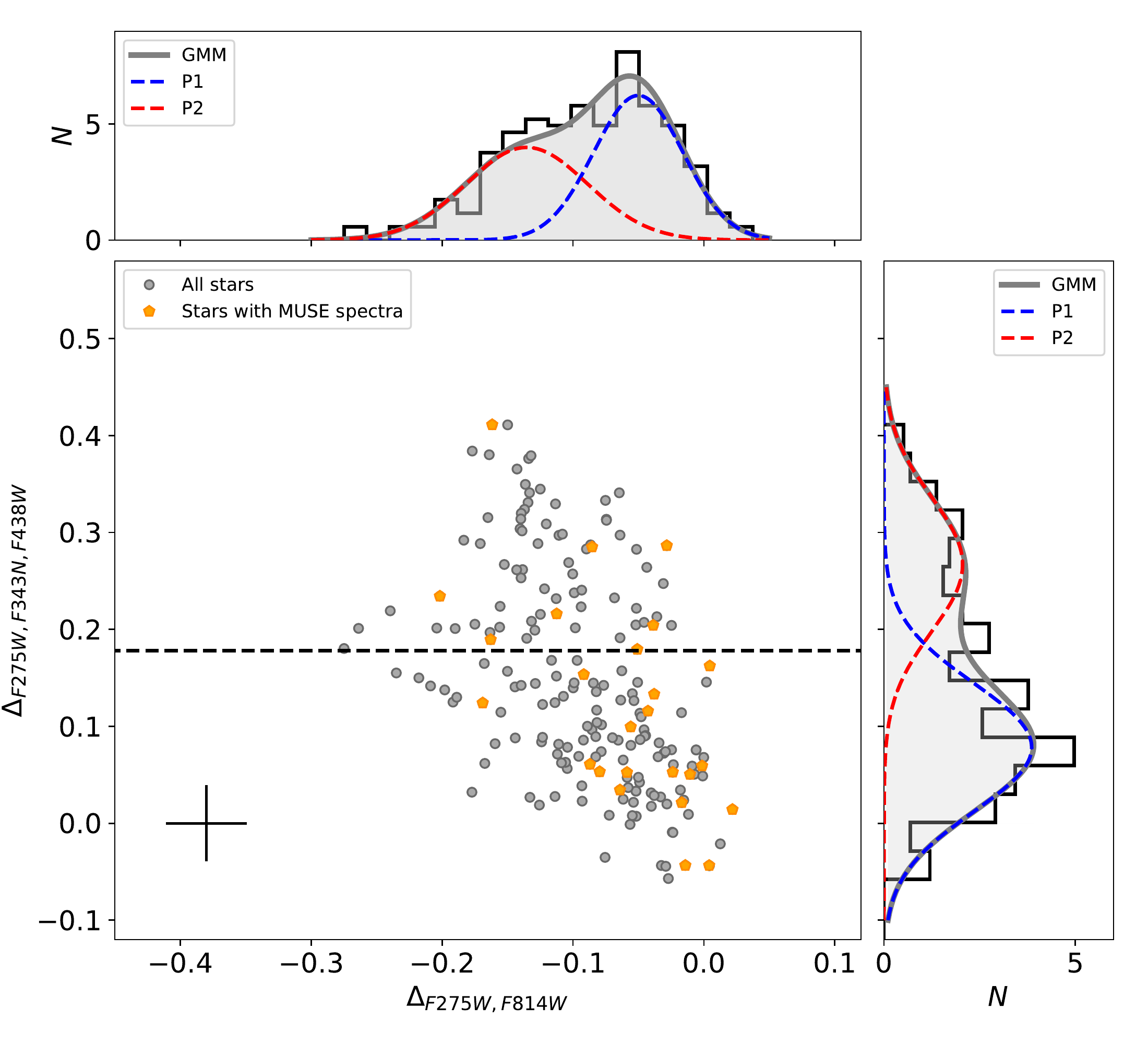}
\caption{Chromosome map of NGC 416 (left) and Lindsay 1 (right). Gray filled circles represent selected RGB stars while orange filled circles indicate stars that have a corresponding MUSE spectrum. The average photometric error is reported on the lower left of the plot. The black dashed line represents the separation limit between the two populations. The upper and right panels represent the histograms of the distribution in $\Delta_{F275W,F814W}$ and $\Delta_{F275W,F343N,F438W}$ colours, respectively. Results from the GMM fitting are also reported, see text for more details. } 
\label{fig:chm}
\end{figure*}

The resulting ChM of the RGB of NGC~416 and Lindsay~1 are shown in Fig. \ref{fig:chm}. The ChM is composed of $\Delta_{F275W,F814W}\equiv\Delta$(\uvi) on the x$-$axis and $\Delta_{F275W,F343N,F438W}\equiv\Delta$(\c) on the y-axis. The $\Delta$ represents the distance in a given colour from defined fiducial lines on the edges of the RGB (one on the red edge, one on the blue edge).
The $\Delta_{F275W,F814W}=0$ and $\Delta_{F275W,F343N,F438W}=0$ are defined for stars on the red edge, such that $\Delta_{F275W,F814W}$ is negative towards the blue while $\Delta_{F275W,F343N,F438W}$ is positive towards the blue. 
For more details on how the ChM and fiducial lines were computed we refer the interested reader to \cite{saracino19,saracino20}. 
MUSE RGB spectroscopic targets are indicated with orange filled circles. 

In the upper and right panels of each ChM of Fig. \ref{fig:chm} we report the histograms of the distributions in $\Delta_{F275W,F814W}$ and $\Delta_{F275W,F343N,F438W}$, respectively. Additionally, we fit the colour distributions with Gaussian Mixture Models (GMMs) to identify the presence of multiple gaussian components. We used the \texttt{SCIKIT-LEARN} python package called \texttt{MIXTURE}\footnote{\url{http://scikit-learn.org/stable/modules/mixture.html}}, which consists of an expectation-maximization algorithm for fitting mixtures of Gaussian models. The best fit to the total population is shown as a solid gray line, while the single components are indicated as dashed blue and red lines for the P1 and P2 populations, respectively. In both pseudo-colours the GMM fit finds two Gaussian components. 

We used the ChM pseudo-colour $\Delta_{F275W,F343N,F438W}$ to separate the two populations with different N content. To this purpose, we separate the populations at $\Delta_{F275W,F343N,F438W} = 0.23$ mag for NGC~416 and $\Delta_{F275W,F343N,F438W} = 0.19$ mag for Lindsay~1, where P1 stars are below this value and P2 stars are above. This value has been chosen to be the point, in the y-axis, where the two gaussian functions cross each other.

The individual spectra are too noisy and the MUSE resolution is too low ($R\sim 3000$) to allow the measurement of elemental abundances. For such a reason, the spectra of each sub-population are combined to obtain a high SNR spectrum from which elemental abundances can be measured in a differential way. 
The strength of such a differential method is that any systematic differences between the synthetic templates and the observed spectra are irrelevant, as long as their effect is comparable for P1 and P2. Having removed the systematic uncertainties, we can look for smaller differences in Na.

We then combined all MUSE spectra by adopting the average and only including stars with SNR$>20$ (see Section \ref{subsec:muse}). For NGC~416 we obtained 76 stars belonging to P1 and 69 stars belonging to P2. The final SNR of the P1 spectrum is $\sim$440, while for P2 is $\sim$400. The fraction of P2 stars with available MUSE spectra is $\sim 47$\% for NGC~416. If we calculate the fraction of total P2 stars (grey circles in Fig. \ref{fig:chm}) by using the same ChM selection, we obtain $\sim 46$\%. Hence, the spectral sample of P2 stars well represents the total population fraction.

\begin{table*}
\caption{Photometric and spectroscopic information of the RGB MUSE targets used for the analysis of NGC~416. Columns represent the following: (1) ID of the star, (2) and (3) right ascension and declination in degrees, (4) and (5) $\Delta$ colours for the ChM in mag, (6) Signal-to-Noise ratio of the star spectrum, (7) metallicity [Fe/H] in dex, (8) radial velocity in ${\rm km/s}$, (9) number of spectra available for each star, (10) population tag based on the ChM selection (Section \ref{subsec:chm}). The full Table will be available in the online version of the paper.}
\begin{tabular}{c c c c c c c c c c}
\hline
SourceID & RA & Dec &  $\Delta_{F275W,F814W}$ & $\Delta_{F275W,F343N,F438W}$ & SNR & $[Fe/H]$ & RV & N$_{exp}$ & Pop \\ 
(1) & (2) & (3) & (4) & (5) & (6) & (7) & (8) & (9) & (10)\\
\hline
1 & 16.9760971 & -72.3558426 & -0.2129 & 0.1853 & 26.73 & -1.16$\pm$0.10 & 157.1$\pm$4.0 & 2& P1\\
2 & 16.9967976 & -72.3575821 & -0.0157 & 0.0544 & 24.68 & -0.91$\pm$0.07 & 157.8$\pm$4.1 & 2& P1\\ 
3 & 16.9963894 & -72.349823  & -0.3282 & 0.1490 & 24.14 & -1.01$\pm$0.12 & 153.2$\pm$5.4 & 2 & P1\\ 
$\dots$ & $\dots$ & $\dots$ & $\dots$ & $\dots$ & $\dots$ & $\dots$ & $\dots$ & $\dots$ & $\dots$ \\
77 & 16.9931889 & -72.3597717 & -0.1708 &  0.3408 & 25.82 & -1.04$\pm$0.11 & 152.8$\pm$ 3.6 & 2 & P2\\
78 & 16.9985065 & -72.3531036 & -0.2397 & 0.3441 & 20.48 & -0.96$\pm$0.13 & 161.2$\pm$5.2 & 2 & P2\\
79 & 17.0074463 & -72.3584366 & -0.1874 & 0.4224 & 23.05 & -1.00$\pm$0.11 & 158.1$\pm$ 4.7 & 2 & P2\\
$\dots$ & $\dots$ & $\dots$ & $\dots$ & $\dots$ & $\dots$ & $\dots$ & $\dots$ & $\dots$ & $\dots$ \\
\hline
\end{tabular}
\label{tab:n416_rgb}
\end{table*}

\begin{figure*}
\centering
\includegraphics[scale=0.4]{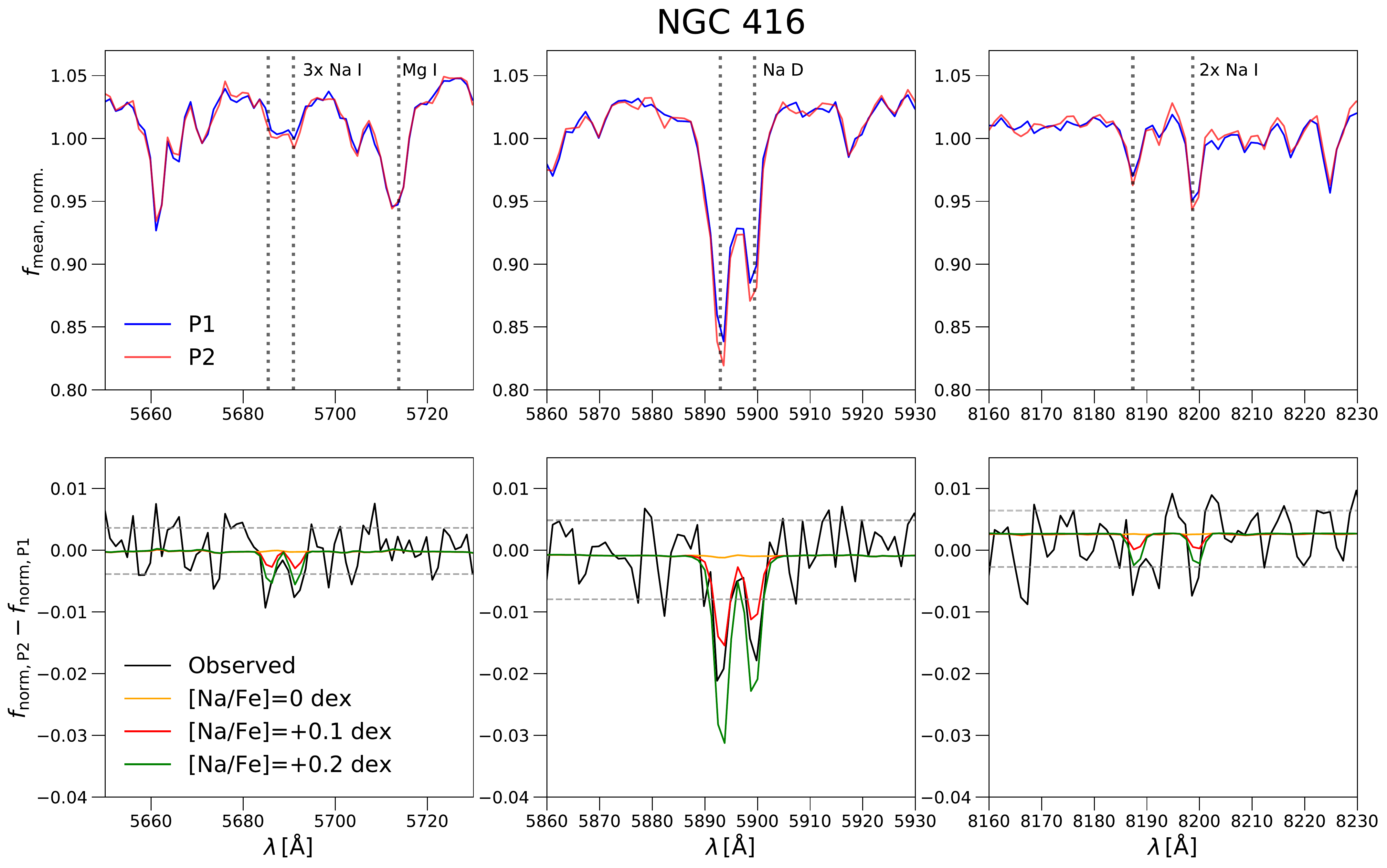}
\caption{\textit{Top Panels:} P1 (P2) MUSE normalised combined spectra for NGC 416 are represented as a solid blue (red) line. Na lines are indicated in each panel. \textit{Bottom Panels:} the flux difference between the observed combined P2 and P1 spectra is indicated with black solid lines. Models taking into account different spreads in Na are indicated with coloured solid lines. The gray dashed lines represent the $1\sigma$ dispersion around the average of the flux difference throughout each wavelength window. See text for more details.} 
\label{fig:n416_rgb}
\end{figure*}

For Lindsay~1, unfortunately, after selecting stars with SNR$>20$, we have only 25 stars left (see \S \ref{subsec:muse}). We combined the MUSE spectra belonging to P1 (19 stars) and P2 (6 stars). For Lindsay~1, the fraction of P2 stars with available MUSE spectra is $\sim 24$\%, while the total P2 fraction is $\sim$34 \%. In this case, we are underestimating the sample of the P2 population, hence the estimated mean Na variation will likely be a lower limit.
The final SNR of the P1 spectrum is $\sim$100, while for P2 is $\sim$30. Also, P1 and P2 stars have different luminosity functions, as selected P2 stars are fainter than the RGB bump, while we have a handful of stars in the P1 which are reaching the tip of the RGB. We made several attempts to improve the statistics in Lindsay~1. We also tried to use lower SNR stars (down to SNR$=10$) and we tried to use the verticalised RGB in $C_{F336W,F438W,F343N}\equiv (m_{F336W}-m_{F438W})-(m_{F438W}-m_{F343N})$ colours \citep{niederhofer17b,martocchia19} to separate the two populations. The results stay unchanged, although the combined spectra are more noisy than our first attempt. We also tried to separate the two populations from the $C_{F275W,F555W,F814W}\equiv (m_{F275W}-m_{F555W})-(m_{F555W}-m_{F814W})$ vs. \i CMD, however the split was not evident enough to make the separation between P1 and P2 trivial. Hence, we decided to keep the combination coming from the ChM. 

Tables \ref{tab:n416_rgb}, \ref{tab:l1} and \ref{tab:n416_hb} report the spectroscopic and photometric information of the NGC~416 RGB, RC and Lindsay~1 RGB MUSE targets, respectively.

In combining the spectra we considered all the selected stars from the ChMs to be members of the cluster. We checked the radial velocity (RV) and metallicity ([Fe/H]) distributions of the RGB stars used in this analysis and we observe no significant spread or outliers. We reported how RVs and [Fe/H] were calculated from MUSE spectra in Section \ref{sec:obs}. The mean RV for NGC~416 is $\sim$156 km/s with a 1$\sigma$ scatter of $5$ km/s. The mean [Fe/H] is $-1.0$ dex (consistent with isochrone fitting estimates, e.g. \citealt{glatt08}) with a 1$\sigma$ scatter of $\sim$0.1 dex. For Lindsay~1 we obtain a mean RV$\sim$141 km/s, $\sigma=4$ km/s (consistent with previous works, e.g. \citealt{hollyhead17}) and a mean [Fe/H]$\sim -1.1$ dex (consistent with previous measurements and isochrone fitting, \citealt{dacosta98,glatt08}), $\sigma \sim$0.1 dex. We note that Lindsay~1 is not expected to be strongly contaminated from field stars as it is located in the outskirts of the SMC \citep{glatt08}.

\begin{table*}
\caption{Photometric and spectroscopic information of the RGB MUSE targets used for the analysis of Lindsay~1. Columns as in Table \ref{tab:n416_rgb}. The full Table will be available in the online version of the paper.}
\begin{tabular}{c c c c c c c c c c}
\hline
SourceID & RA & Dec &  $\Delta_{F275W,F814W}$ & $\Delta_{F275W,F343N,F438W}$ & SNR & $[Fe/H]$ & RV & N$_{exp}$ & Pop \\  
(1) & (2) & (3) & (4) & (5) & (6) & (7) & (8) & (9) & (10)\\
\hline
1 & 0.9913839 & -73.4699327 & 0.0218 &  0.0142 & 45.43 & -1.04$\pm$0.02 & 139.1$\pm$0.8 & 4 & P1\\
2 & 0.9864344 & -73.4718961 & -0.0106 & 0.0505 & 31.15 & -1.05$\pm$0.11 & 138.4$\pm$3.8 & 4 & P1\\
3 & 0.9713206 & -73.4778053 & -0.0238 & 0.0527 & 20.28 & -1.18$\pm$0.15 & 144.1$\pm$5.9 & 4  & P1\\
$\dots$ & $\dots$ & $\dots$ & $\dots$ & $\dots$ & $\dots$ & $\dots$ & $\dots$ & $\dots$ & $\dots$\\
21 & 0.9712204 & -73.4759099 & -0.2017 &  0.2342 & 36.10 & -1.06$\pm$0.05 & 143.8$\pm$2.0 &   4 &P2\\      
22 & 0.9724604 & -73.4746504 & -0.0284 & 0.2864 &  36.61 & -1.08$\pm$0.05 &  144.7$\pm$2.1 &   4 &P2\\
23 & 0.9475757 & -73.4713739 & -0.1125 & 0.2160 &  30.30 & -0.94$\pm$0.08 &  141.3$\pm$ 3.6 &  4  &  P2\\
$\dots$ & $\dots$ & $\dots$ & $\dots$ & $\dots$ & $\dots$ & $\dots$ & $\dots$ & $\dots$ & $\dots$ \\
\hline
\end{tabular}
\label{tab:l1}
\end{table*}

\begin{figure*}
\centering
\includegraphics[scale=0.4]{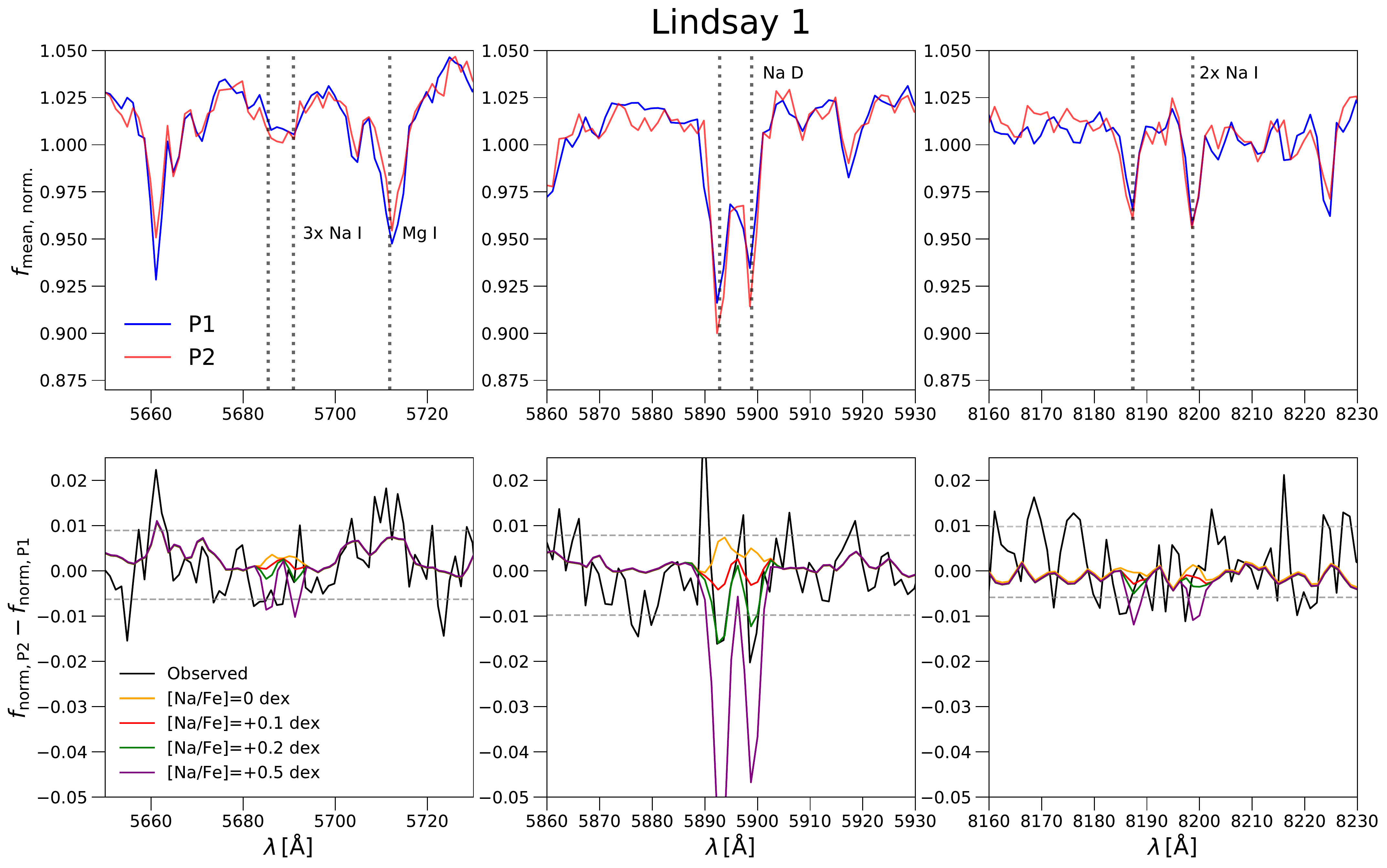}
\caption{\textit{Top Panels:} P1 (P2) MUSE normalised combined spectra for Lindsay~1 are represented as a solid blue (red) line. Na lines are indicated in each panel. \textit{Bottom Panels:} the flux difference between the observed combined P2 and P1 spectra is indicated with black solid lines. Models taking into account different spreads in Na are indicated with coloured solid lines. See text for more details.
} 
\label{fig:l1_rgb}
\end{figure*}

\section{Results}\label{sec:res}
We combined all MUSE spectra belonging to each of the two identified sub-populations for both clusters, as reported in Section \ref{sec:an}. The main goal of this analysis is to see whether there are differences in the mean abundances between the populations within each cluster, which should manifest as differences in specific spectral lines. For this, we followed the same approach as the analysis from \cite{latour19}. 

Given the small expected and measured velocity dispersion and the resolution of MUSE, we did not shift the spectra for radial velocities.
The top panels of Figure \ref{fig:n416_rgb} shows the comparison between the P1 (blue line) and P2 (red line) combined spectra around several Na lines for NGC~416. The same is reported in Fig. \ref{fig:l1_rgb} for Lindsay~1. The bottom panels show the difference between the normalised P2 and P1 spectra as black solid lines. The gray dashed lines in the bottom panels represent the $1\sigma$ dispersion around the average of the flux difference throughout each wavelength window. 
We observe a significant difference in the NaD lines, but no difference in the other Na I lines (left and right top panels of Fig. \ref{fig:n416_rgb}). 
For a list of lines we refer to \citeauthor{latour19} (\citeyear{latour19}, their Table~2). 

From this comparison, it is clear that there is a difference between the two spectra around the NaD lines (5889.95, 5895.92\AA) for NGC~416, with the P2 having deeper lines. 
This implies that the P2 population has an enhanced Na abundance with respect to the P1 population. We also observe a small difference in the lines at 5682.63, 5688.19, 5688.21\AA. This is not statistically significant, however it is consistent with P2 being slightly enhanced in Na with respect to P1.

For Lindsay~1, we observe no difference for the Na I lines at $\sim$5700\AA \, and $\sim$8200\AA. However, despite the lower SNR of Lindsay~1 spectra (combination of 19 and 6 stars for P1 and P2, respectively, Section \ref{sec:an}), we observe a difference in the NaD lines. 
We conclude that there is also a difference in Na between the two populations in Lindsay~1. 

To quantify the detections, we fit a double Gaussian to the data reported in the bottom central panel of Figs. \ref{fig:n416_rgb} and \ref{fig:l1_rgb}, such that the observed difference in the NaD lines is converted into an equivalent width (EW). A common full-width at half maximum of $2.5\,{\text \AA}$ was adopted, since it corresponds to the spectral resolution of MUSE. The areas of the Gaussians were calculated through a linear least square fit. Additionally, we estimated the uncertainty on the areas by adding random noise to the observed flux difference, according to the grey horizontal dashed line shown in Figs. \ref{fig:n416_rgb} and \ref{fig:l1_rgb}. 
This is done 100 times and each time we calculate the EW. The error on the EW is then the standard deviation of such 100 EWs.
We obtain a difference of $-117\pm21\,{\rm m}{\text \AA}$ for NGC~416 ($\sim$5.6$\sigma$) and $-93\pm28\,{\rm m}{\text \AA}$ for Lindsay~1 ($\sim$3.3$\sigma$). 

We repeated this analysis by using a different population selection approach. We used a two-dimensional Gaussian Mixture Model on the Chromosome map (Fig. \ref{fig:chm}) with the same \texttt{SCIKIT-LEARN} python package called \texttt{MIXTURE}, however this time we considered both axes of the ChM to assign a star to each population. The main differences with respect to our preferred method (see Section \ref{sec:an}) are a handful of stars near the border between the two populations. We obtain the exact same result in terms of significance of the detection for NGC~416, i.e. EW$=-115\pm21\,{\rm m}{\text \AA}$. For Lindsay~1, we still observe a difference in the NaD lines. However, given the difference in the luminosity functions of the populations and that we have very few stars in the P2, we obtain $-36\pm28\,{\rm m}{\text \AA}$, lowering the significance to $\sim 1.3\sigma$.

As differential reddening (DR) may not be negligible in NGC~416, we also investigated whether it was affecting our results. However, since we are combining a relatively large number of spectra, from various spatial locations around the cluster, the effect of DR should effectively average out. We found that the mean $\delta E(B-V)$ for P1 and P2 is not significantly different, being $\delta E(B-V)_{\rm mean, P1}\sim$-0.005 mag and $\delta E(B-V)_{\rm mean, P2}\sim$-0.01 mag, hence no significant effect on the P1/P2 difference is expected. 

As an additional test, we only combined P1 and P2 stars that have positive $\delta E(B-V)$ values and also only P1 and P2 stars that have negative $\delta E(B-V)$ values (in these cases, the mean $\delta E(B-V)$ is the same for both populations). The difference between the NaD lines stays unchanged in both cases, with P2 always showing deeper NaD lines. 

Finally, we also checked whether the observed difference in the NaD lines might be affected by the presence of the brightest stars in the sample. The same analysis reported in this Section and in Section \ref{subsec:comp} was performed by stacking only faint spectra, i.e. stars with \i$>$18 mag for both clusters. The EWs and $\Delta$[Na/Fe] are consistent to what has been obtained before, hence we are confident that the results reported in this work are not dependent on the inclusion of bright stars in the final combined spectra.

\subsection{Synthetic Models}\label{subsec:models}

Synthetic spectra and photometry were calculated in order to compare theoretical models to the data (see \S \ref{subsec:comp}). For the details about the computation of the models we refer the reader to \cite{martocchia17}.
For NGC~416 we used a $\sim$6.3 Gyr, [Fe/H]$ = -1.0$ dex \citep{glatt08} MIST isochrone \citep[version 1.2][]{2011ApJS..192....3P, 2016ApJS..222....8D, 2016ApJ...823..102C}.
We selected 35 evolution points evenly spaced in $\log T_{\rm eff}$-$\log L$ space between the start of the sub giant branch and the tip of the RGB.
For each of these points we computed a model atmosphere using \textsc{ATLAS12} \citep{1970SAOSR.309.....K, 2005MSAIS...8...14K} before using \textsc{SYNTHE} \citep{1979SAOSR.387.....K, 1981SAOSR.391.....K} to synthesize a spectrum for each.
We obtained synthetic photometry from each model spectrum by using the filter curves and zeropoints provided by the WFC3 website.
We calculated these models for five different chemical mixtures of light element abundances assuming the \citet{2009ARA&A..47..481A} solar abundance scale.
First, we computed a model with scaled solar abundances  ([C/Fe] = [N/Fe] = [O/Fe] = [Na/Fe] $= 0$), then a model with enhanced N and depleted C and O but scaled solar Na ([N/Fe] $= +0.5$, [C/Fe] = [O/Fe] $= -0.1$, [Na/Fe] = $0$).
Finally, we calculated models with the same enhanced N and depleted C and O but three different Na enhancements ([Na/Fe] $= +0.1$, [Na/Fe] $= +0.2$, and [Na/Fe] $= +0.5$). For our comparison between models and observations to infer the Na variation present in each cluster (see next Section \ref{subsec:comp}), we used a N-enhanced model to describe both P1 and P2. This choice is motivated by the effect of the first dredge up on RGB stars at such younger ages \citep{salaris20}. We note that, as an additional test, we computed the Na variations also by assuming solar scaled abundances only for the P1, and we obtained the exact same result as in the first case. Hence, variations of C, N, O do not affect the measurement of the Na difference.
For all of the models we kept the He abundance and all other element abundances constant at their scaled solar values.

We carried out the same procedure to simulate the stellar population within Lindsay~1.  The only difference to the above is that we adopted an isochrone with an age of $\sim 8$~Gyr \citep{glatt08} and a metallicity of [Fe/H]$=-1.1$ dex.

For each observed MUSE spectrum, the closest template model that matched the F814W magnitude was chosen and then all template models for the selected chemical mixture were combined in the same way as the observed spectra. 

\subsection{Comparison with models}\label{subsec:comp}
In order to estimate the sodium variation between P1 and P2, we compared the observed flux difference with the models from \S \ref{subsec:models}. Note that we are estimating the mean Na variation between the two populations and not its maximum range of extension. This is shown for NGC~416 in the bottom panels of Fig. \ref{fig:n416_rgb}, and for Lindsay~1 in the bottom panels of Fig. \ref{fig:l1_rgb}.  In Figs. \ref{fig:n416_rgb} we report models that have [Na/Fe]$=0$ dex, [Na/Fe]$=+0.1$ dex, [Na/Fe]$=+0.2$ dex as yellow, red and green solid lines, respectively, while in Fig. \ref{fig:l1_rgb} the purple lines represent a model with [Na/Fe]$=+0.5$ dex. From the comparison around the NaD lines, it is clear that a difference in sodium $+0.1<\Delta$[Na/Fe]$<+0.2$ dex is needed to explain the difference in the NaD lines between the two populations of NGC 416 (central bottom panel). By comparing theoretical models with the observed difference between the two populations we obtain $\Delta$[Na/Fe]$=0.18\pm0.04$ dex, a $\sim$4.5$\sigma$ detection for NGC~416. This was calculated by assuming that the line strength EW increases linearly with the [Na/Fe] variation. 
By applying the same methodology to the observations and models of Lindsay~1, we obtain $\Delta$[Na/Fe]$=0.24\pm0.05$ dex, which represents a high detection, a $\sim$5$\sigma$ detection.
The error on the [Na/Fe] variation was calculated by adding noise to the synthetic spectra. We added the noise with the method reported in Section \ref{sec:res}, based on the observed errors, i.e. the grey dashed horizontal lines in the bottom panels of Fig. \ref{fig:n416_rgb} and \ref{fig:l1_rgb}. 
More specifically, we added Gaussian noise to the model flux difference for 100 times and each time we calculated the EW. The error on the model EW is then the standard deviation of the 100 EWs, as done with the data (see Section \ref{sec:res}). Finally, since we assume that the EW scales linearly with the Na difference, we estimated the error on $\Delta$[Na/Fe] by means of the error propagation. 

We note that the errors on the mean Na variation of Lindsay 1 are, most likely, considerably larger if we take into account the lower SNR of the spectra, the small number of stars involved and the effect of the selection (see Section \ref{sec:res}). \footnote{Additionally, we estimated the errors on the EW (see Section \ref{sec:res}) via a bootstrap technique. Briefly, we combined the P1 and P2 spectra by leaving one random spectrum out from the sample and replacing it with another random one from the other population. We repeated this 1,000 times and for each time we calculated the EW of the NaD lines in the observed flux difference, in the same way we did in Section \ref{sec:res}. We used the standard deviation of the EW distribution for a new estimate of the error. When comparing to models for the $\Delta$[Na/Fe] computation we obtain still that the Na variation is $\sim$4$\sigma$ significant for NGC~416, while it is a $\sim$4.4$\sigma$ detection for Lindsay~1.}

The NaD lines are useful diagnostic of Na abundances at low resolution but 
they are strong/damped lines sensitive to the photospheric velocity 
fields. In fact, the depth of the NaD lines varies as a function of the microturbulent velocity ($\xi$).
In our calculation of the synthetic spectra, we assumed a microturbulent velocity that varies with the surface gravity log $g$. For stars with log $g<1$, we used $\xi$=2 km/s, while for $1<$log $g<4$ we adopted the formula $\xi$=(-log $g$+7)$/3$ (this is a linear fit between the values we adopt above log g$=$1 and below log $g=$4). Finally, for log $g>4$, we used $\xi$=1 km/s. If we assumed that the observed flux difference would be due to differences in $\xi$, it would imply that on average P1 and P2 would have different microturbulent velocities. Since the microturbulent velocity depends on the temperature and gravity of stars, this would imply very different parameters for the different populations. Additionally, we demonstrated above that the inclusion of bright stars does not change the results presented in the paper (Section \ref{sec:res}). We do not see any reason for the average $\xi$ of the populations to differ in a significant way.

We estimated the average microturbulent velocity of P1 and P2 from the synthetic spectra, if we assume that the $\Delta$[Na/Fe] is +0.2 dex, for both clusters. As explained at the end of Section \ref{subsec:models}, for each observed spectrum in the sample, we created a synthetic one, i.e. the closest template model that matched the F814W magnitude.
Hence, we assigned a log $g$ to each star and we estimated what is the expected average $v_{mic}$ of each population. We find that the microturbulent velocities of P1 and P2 are the same to the second decimal digit for both clusters. We are confident that the microturbulent velocity is not playing any role in the flux difference we observe in the NaD lines.

\begin{figure}
\centering
\includegraphics[scale=0.5]{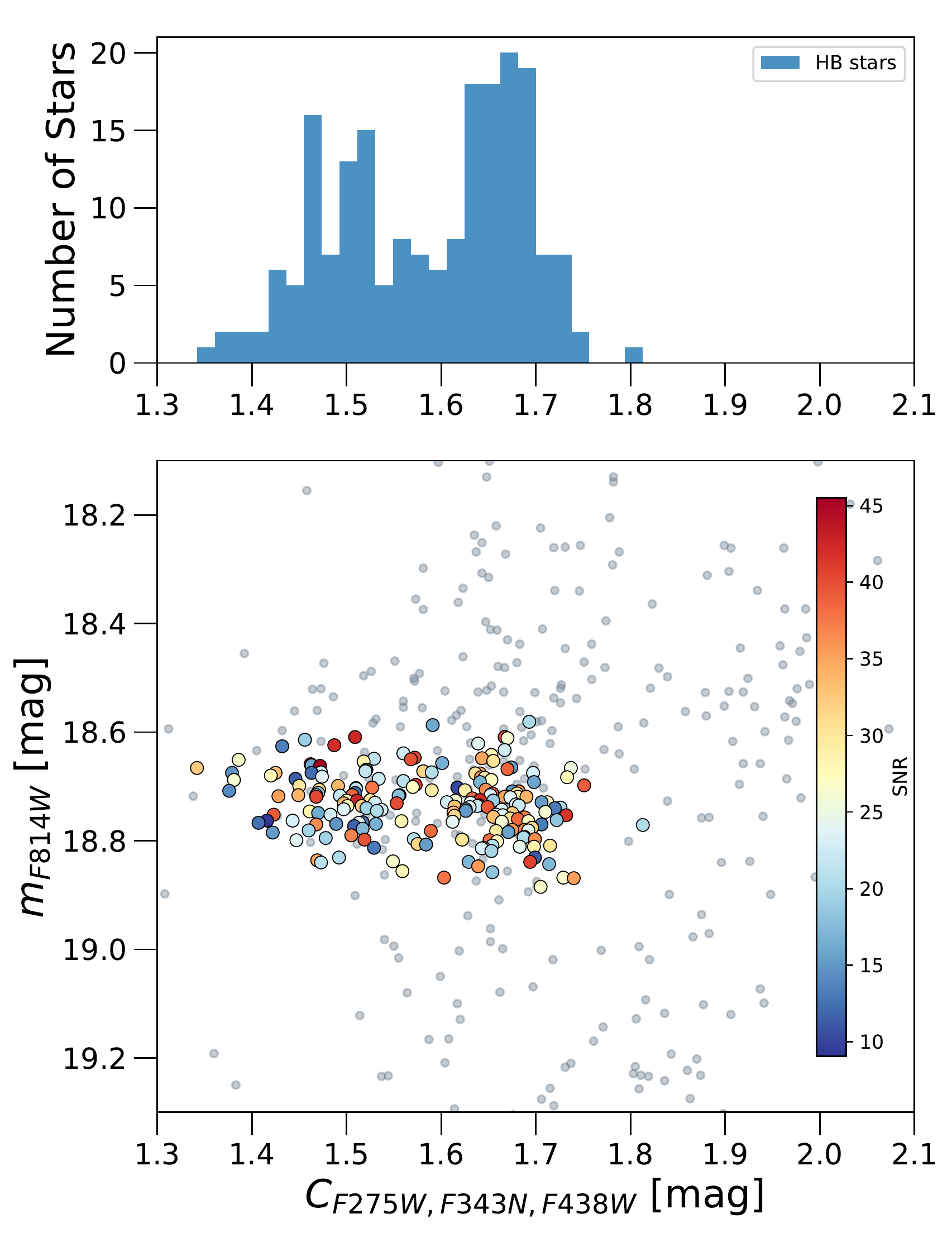}
\caption{\textit{Bottom panel}: \c vs. \i CMD of NGC 416 zoomed in the RC region, where MUSE RC targets are colour-coded by their SNR. \textit{Upper panel}: histogram of the distribution of RC stars in \c colours.} 
\label{fig:hbcmd}
\end{figure}

\section{The double Red Clump of NGC~416}\label{sec:hbs}
We selected red clump (RC) helium core burning stars in NGC 416 from the optical CMD \vi vs. \i and then plotted them in the \c vs. \i CMD.
We noticed that in such a pseudo-colour (\c), NGC~416 shows a bimodal red clump. 
This is shown in the bottom panel of Fig. \ref{fig:hbcmd} where the CMD is zoomed in on the RC region. MUSE spectroscopic targets for the RC are colour-coded by their SNR. For a better visual representation, the upper panel shows the histogram of the distribution of the RC stars in \c colours with a bin size of 0.02 mag. We then separated the two populations of the RC, by creating a ChM of the RC in the same way as it has been done for the RGB (see Section \ref{subsec:chm}).
Two fiducial lines have been defined at the edges of the RC in both the colour \uvi and pseudo-colour \c. They have been normalized as done for the RGB stars. A slightly different normalisation would only affect the extension of both axis, but it would not change anything in a relative sense. The ChM of the RC is only used here to facilitate the separation between the two populations and we do not intend to use it for a one-to-one comparison with other ChMs.

\begin{figure}
\centering
\includegraphics[scale=0.4]{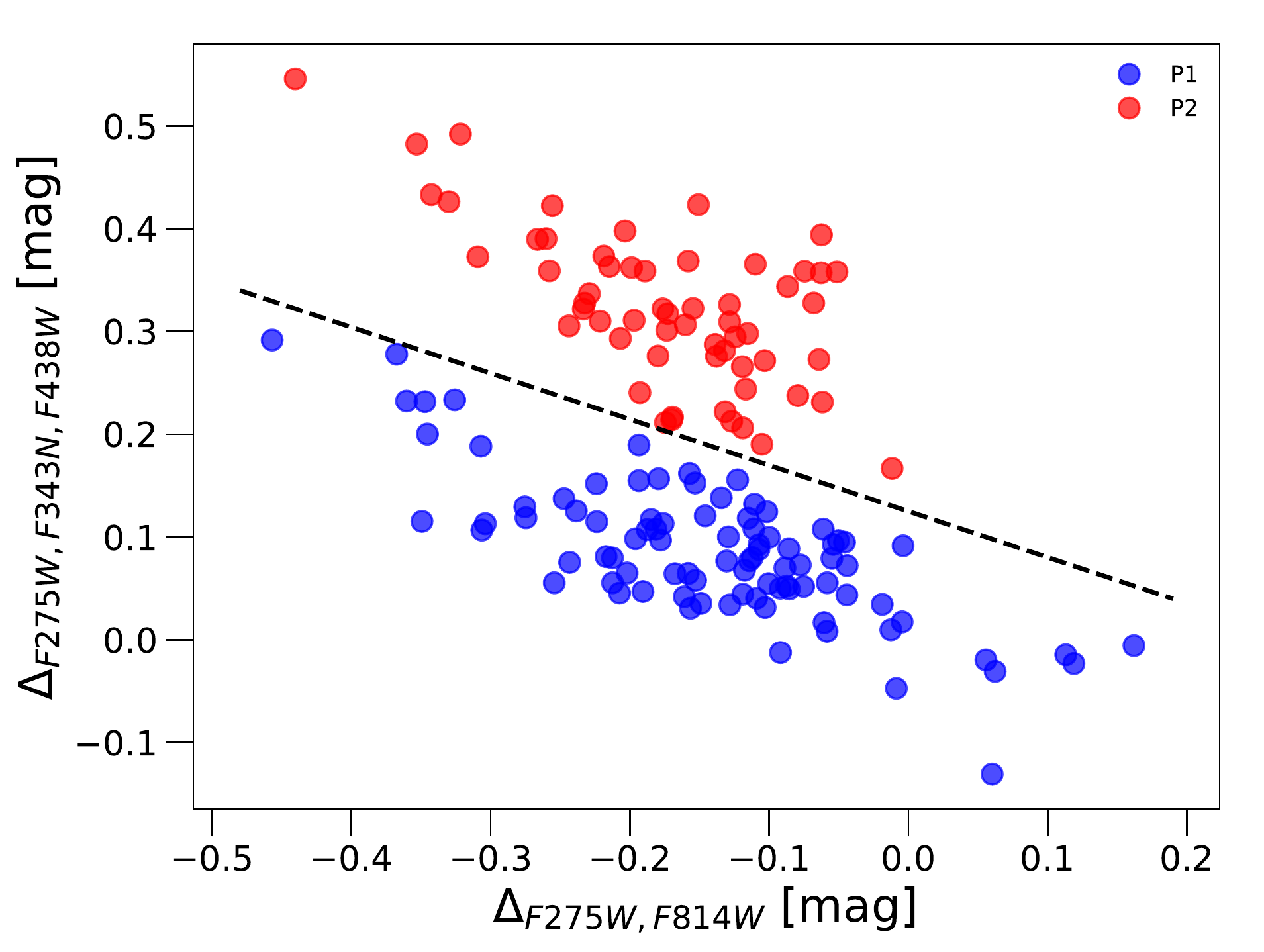}
\caption{Chromosome map of NGC 416 RC stars. Blue (red) filled circles represent selected RC P1 (P2) stars.} 
\label{fig:hbchm}
\end{figure}

The ChM of the RC of NGC~416 is shown in Fig. \ref{fig:hbchm}, where we indicate the selection of P1 (P2) stars with blue (red) filled circles. The selection was made with the dashed black line shown in Fig. \ref{fig:hbchm}, following the approach used in \cite{milone17} to distinguish P1 and P2 stars. We also made a selection by simply considering a horizontal line (as in Fig. \ref{fig:chm} for the RGB) but our results do not change significantly. 

\begin{figure}
\centering
\includegraphics[scale=0.4]{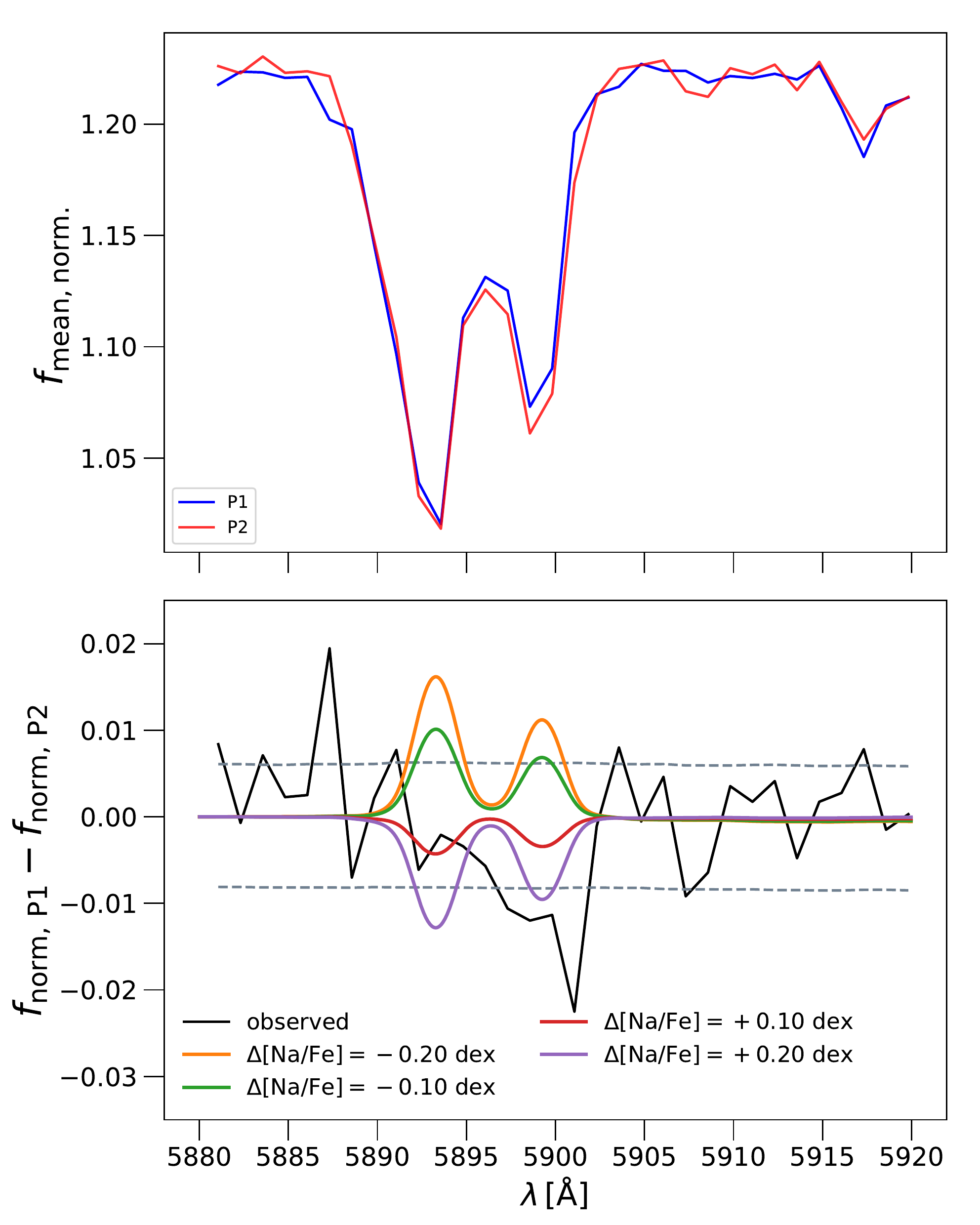}
\caption{\textit{Top Panel:} Combined spectra of NGC 416 RC stars for the P1 (P2) population indicated as a blue (red) curve, zoomed in around the NaD lines. \textit{Bottom panel:} difference between P2 and P1 spectra as a function of wavelength. The dashed gray lines represent one standard deviation dispersion from the mean line. Models taking into account different spreads in Na are indicated with coloured solid lines. See text for more details.} 
\label{fig:hb}
\end{figure}

We then combined all RC MUSE spectra belonging to P1 (90 stars) and P2 (59 stars), as was done for the RGB. In this case, the fraction of P2 stars over the total is $\sim$ 40\%, slightly lower than for the RGB sample. We show the comparison between the two spectra around the NaD lines in Fig. \ref{fig:hb}. In the bottom panel we report the observed difference in flux between P2 and P1. There is a hint of difference in the reddest NaD line between the two populations, but this is only significant at less than $3\sigma$. Indeed, we found a difference in the NaD lines of the RC of NGC~416 of $-44\pm16\,{\rm m}{\text \AA}$.
To estimate the Na variation in the RC between the P1 and P2 we used the exact same method reported in Section \ref{subsec:comp} for the RGB. We compared the observed flux difference between P1 and P2 with models, which are reported in the bottom panel of Fig. \ref{fig:hb}. Models for the RC were computed as in \cite{latour19}. We obtain a Na variation  $\Delta$[Na/Fe]$=0.15\pm0.05$ dex, with a significance of 3$\sigma$. This is less significant than the RGB result, however the results are consistent. 
Table \ref{tab:n416_hb} reports the photometric and spectroscopic information of the NGC~416 RC targets used in this paper. 

In the RC, we noticed that the flux difference only appears on one NaD line, i.e. the red one. Hence, we made a bootstrap test to check if the difference observed on the red line was only due to noise. The bootstrap was repeated for 1000 times. For each time, 5 random stars were removed from one population and combined with the other one. We compared this bootstrapped populations with the synthetic models as above so that we obtain a $\Delta$[Na/Fe] each time. The difference in the reddest Na line is observed most of the times and sometimes it is observed also in the bluest line, however this is much fainter. The exact reason why this happens is unfortunately unknown and futher work is needed. We averaged over the 1000 $\Delta$[Na/Fe] values obtained  from the bootstrap and we obtained $\Delta$[Na/Fe](mean)$=$0.15 dex with a standard deviation of 0.07 dex. Hence, we are confident that the flux difference observed is not due to noise.

For Lindsay~1, we did not find any clear separation for the RC, hence we could not carry out a similar analysis there.

\begin{table*}
\caption{Photometric and spectroscopic information of the RC MUSE targets used for the analysis of NGC~416. Columns as in Table \ref{tab:n416_rgb}. The full Table will be available in the online version of the paper.}
\begin{tabular}{c c c c c c c c c c}
\hline
SourceID & RA & Dec &  $\Delta_{F275W,F814W}$ & $\Delta_{F275W,F343N,F438W}$ & SNR & $[Fe/H]$ & RV & N$_{exp}$ & Pop \\
(1) & (2) & (3) & (4) & (5) & (6) & (7) & (8) & (9) & (10)\\
\hline
1 & 17.0005951 & -72.3543701 & -0.0442 & 0.0437 &  20.14 & -1.18$\pm$0.11 & 148.8$\pm$4.2 & 2 & P1\\ 
2 &  16.9880066 & -72.3499756 & -0.1121 & 0.0801 &  39.84 & -1.01$\pm$0.06 &   164.9$\pm$3.3 & 2 & P1\\  
3 & 16.9997978 & -72.3521118 & -0.0777 & 0.0729 &  25.86 & -1.03$\pm$0.08 &   164.6$\pm$2.7 & 2 & P1\\
$\dots$ & $\dots$ & $\dots$ & $\dots$ & $\dots$ & $\dots$ & $\dots$ & $\dots$ & $\dots$ & $\dots$ \\
91 &  16.9858017 & -72.3497162 & -0.1284 & 0.3093 & 42.65 & -1.09$\pm$0.05 &   164.2$\pm$2.4 & 2 & P2\\  
92 &  16.9906654 & -72.3507614 & -0.2292 & 0.3369 & 42.18 & -1.05$\pm$0.05 &   158.6$\pm$2.3 & 2 & P2\\  
93 & 17.0103226 & -72.3531113 & -0.1699 & 0.2142 &  40.11 & -1.04$\pm$0.04 &    162.3$\pm$2.1 & 2 &  P2\\  
$\dots$ & $\dots$ & $\dots$ & $\dots$ & $\dots$ & $\dots$ & $\dots$ & $\dots$ & $\dots$ & $\dots$ \\
\hline
\end{tabular}
\label{tab:n416_hb}
\end{table*}

\section{Discussion and conclusions}\label{sec:disc}
In this paper we used a combination of photometric and spectroscopic techniques to search for mean differences in elemental abundances in 
two intermediate age massive clusters in the SMC, namely NGC~416 ($\sim$6.5 Gyr old, \citealt{glatt08}) and Lindsay~1 ($\sim$7.5 Gyr old, \citealt{glatt08}). By following the approach in \cite{latour19}, we first used the available HST photometry and chromosome map of each cluster to separate the N-rich and N-poor populations on the RGB (Section \ref{sec:an}). Next, we combined the available MUSE spectra of the individual stars for each population (Section \ref{sec:res}) and used the difference spectrum to search for differences in Na. We observed a significant difference in the NaD lines between each population in both clusters. 
We compared the observed difference between the P2 and P1 fluxes with theoretical models specifically calculated with different chemical mixtures (enhanced in N and Na, see Section \ref{subsec:models}). For the RGB of NGC~416, we find a mean difference in sodium between the two populations of $\Delta$[Na/Fe]$=0.18\pm0.04$ dex, while for Lindsay~1 we find a difference of $\Delta$[Na/Fe]$=0.24\pm0.05$ dex.

Additionally, from HST photometry in specific UV filters, we report for the first time a double RC in the cluster NGC~416, representative of two populations with different N content. By combining MUSE spectra of the individual stars in each population, and comparing with models as done in the RGB, we find a difference in Na between the P1 and P2 of $\Delta$[Na/Fe]$=0.15\pm0.05$ dex, which is significant at 3$\sigma$ confidence level. We observe the same sodium variation both in the RGB and RC, enforcing the robustness of our results. 

In the first paper of this series, we also found Na variations in the $\sim$2~Gyr old massive cluster NGC 1978 in the Large Magellanic Cloud (LMC, \citealt{saracino20b}). In this cluster, we found a mean variation $\Delta$[Na/Fe]$\sim0.07 \pm 0.01$ dex, which is slightly smaller than that found for the two intermediate age clusters studied in this work. This is the first time that a difference in Na between the populations are reported for such young clusters, where so far only N variations had been found \citep{niederhofer17b,hollyhead17,martocchia18a}.

Since all three clusters have similar masses ($\sim 2 \times 10^5$ \msun , \citealt{mclaughlin05}), and a correlation between N abundance spreads and cluster age is observed \citep{martocchia19}, it is reasonable to think that the smaller Na difference observed in the younger cluster might be due to an age effect, although this interpretation needs to be confirmed with a larger sample. It will be crucial to establish whether a correlation between Na spread and cluster age exists, as Na is not affected by mixing effects such as N \citep{salaris20}. Follow-up studies with future facilities will also be necessary to confirm the presence or lack of other light elements variations in young star clusters, such as O and C.

A comparison can be performed with the work by \cite{latour19}, where they reported a similar analysis for the old GC NGC~2808 in the MW. 
If we compare our results with the one from \cite{latour19}, we notice that the [Na/Fe] difference obtained in our intermediate age clusters is smaller than the one in NGC~2808, which shows a maximum spread of $\Delta$[Na/Fe]$=0.5\pm0.06$ dex. However, NGC~2808 is a rather extreme Galactic GCs, hosting at least five different populations \citep{milone15b}, and containing some of the most extreme abundance variations found. If we only consider the less extreme populations in NGC 2808 (P1 and P2 in the \citealt{latour19} notation), we see that the variation in Na is quite similar to our results (see Table 3 in \citealt{latour19}), where the variation between P1 and P2 in NGC~2808 is $\Delta$[Na/Fe]$=0.15\pm0.06$ dex. 

Additionally, we compared our results with the difference in [Na/Fe] abundances between P1 and P2 in some Galactic GCs reported in \citeauthor{marino19} (\citeyear{marino19}, see their Table 2). We only made the comparison among clusters with similar metallicities to NGC~416 and Lindsay~1. 
In general, we found that the $\Delta$[Na/Fe] is larger for the Galactic GCs, being $\Delta$[Na/Fe]$\simeq$0.3-0.4 dex for clusters like NGC~6362 ([Fe/H]$\simeq -1.0$ dex) and NGC~5272 ([Fe/H]$\simeq -1.3$ dex). 
Interestingly, this might point towards an age effect, as Na spreads are not affected by the first dredge up. However, 
whether this difference depends on a property of the clusters, i.e. age, or mass of the cluster or even the environment in which it was formed, needs to be further investigated with a larger sample.  

The presence of Na variation in massive young and intermediate age clusters (NGC~1978 from \citealt{saracino20b}, Lindsay~1 and NGC~416 from this study) demonstrates that the MP phenomenon is not only restricted to N but is also seen in Na, just as in the ancient GCs. These results lend further support to the idea that young and 
old massive star clusters are the same objects, just seen at different stages of their lives \cite[e.g.,][]{kruijssen15}. This has important implications; first, young star clusters can then be used to place constraints on MP formation scenarios. 
Second, as globular clusters do not need special conditions in which to form, they can be used in place of much more faint underlying stars as dynamic and stellar population tracers for the formation and evolution of galaxies (e.g., \citealt{forbes18}). 

Some of the important constraints for the origin of MPs and clusters obtained through studies of young massive clusters (YMCs) are: 1) special conditions present in the early Universe are not required for their formation; 2) multiple epochs of star-formation within YMCs are not observed, suggesting that the formation of multiple generations of stars is not possible, or at least is not common (\citealt{cabreraziri14,cabrera_ziri15}), other than in nuclear star clusters, \citep{neumayer20}; 3) that any age difference between the multiple populations is $<20$~Myr \citep[][]{martocchia18b, saracino20}; and 4) that the chemical abundance variations are only observed in stars less massive than $\sim1.5$~\msun, i.e. they do not appear to be present (at least not detectable with current methods) in more massive stars \citep[][]{cabreraziri16}.

\section*{Data Availability}
The data underlying this article are available in the article and in its online supplementary material.

\section*{Acknowledgements}
We thank the referee for constructive comments that helped strengthen the paper.
We gratefully acknowledge Florian Niederhofer for helpful comments and discussion. SM, SS, SK and NB gratefully acknowledge financial support from the European Research Council (ERC-CoG-646928, Multi-Pop). NB also acknowledges support from the Royal Society (University Research Fellowship). 
This work is the part of HST GO-14069 and GO-15630  programs awarded by  The Space Telescope Science Institute, which is operated by the Association of Universities for Research in Astronomy, Inc., for
NASA, under contract NAS5-26555. 
Based on observations collected at the European Southern Observatory under ESO programme 0104.D-0257.
Support for this work was provided by NASA through Hubble Fellowship grant HST-HF2-51387.001-A awarded by the Space Telescope Science Institute, which is operated by the Association of Universities for Research in Astronomy, Inc., for NASA, under contract NAS5-26555.  




\bibliographystyle{mnras}
\bibliography{n416} 

\begin{thebibliography}{}
\makeatletter
\relax
\def\mn@urlcharsother{\let\do\@makeother \do\$\do\&\do\#\do\^\do\_\do\%\do\~}
\def\mn@doi{\begingroup\mn@urlcharsother \@ifnextchar [ {\mn@doi@}
  {\mn@doi@[]}}
\def\mn@doi@[#1]#2{\def\@tempa{#1}\ifx\@tempa\@empty \href
  {http://dx.doi.org/#2} {doi:#2}\else \href {http://dx.doi.org/#2} {#1}\fi
  \endgroup}
\def\mn@eprint#1#2{\mn@eprint@#1:#2::\@nil}
\def\mn@eprint@arXiv#1{\href {http://arxiv.org/abs/#1} {{\tt arXiv:#1}}}
\def\mn@eprint@dblp#1{\href {http://dblp.uni-trier.de/rec/bibtex/#1.xml}
  {dblp:#1}}
\def\mn@eprint@#1:#2:#3:#4\@nil{\def\@tempa {#1}\def\@tempb {#2}\def\@tempc
  {#3}\ifx \@tempc \@empty \let \@tempc \@tempb \let \@tempb \@tempa \fi \ifx
  \@tempb \@empty \def\@tempb {arXiv}\fi \@ifundefined
  {mn@eprint@\@tempb}{\@tempb:\@tempc}{\expandafter \expandafter \csname
  mn@eprint@\@tempb\endcsname \expandafter{\@tempc}}}

\bibitem[\protect\citeauthoryear{{Asplund}, {Grevesse}, {Sauval}  \&
  {Scott}}{{Asplund} et~al.}{2009}]{2009ARA&A..47..481A}
{Asplund} M.,  {Grevesse} N.,  {Sauval} A.~J.,   {Scott} P.,  2009, \mn@doi
  [\araa] {10.1146/annurev.astro.46.060407.145222}, \href
  {https://ui.adsabs.harvard.edu/abs/2009ARA&A..47..481A} {47, 481}

\bibitem[\protect\citeauthoryear{{Bastian} \& {Lardo}}{{Bastian} \&
  {Lardo}}{2015}]{bastian_lardo15}
{Bastian} N.,  {Lardo} C.,  2015, \mn@doi [\mnras] {10.1093/mnras/stv1661},
  \href {https://ui.adsabs.harvard.edu/abs/2015MNRAS.453..357B} {453, 357}

\bibitem[\protect\citeauthoryear{{Bastian} \& {Lardo}}{{Bastian} \&
  {Lardo}}{2018}]{bastianlardo18}
{Bastian} N.,  {Lardo} C.,  2018, \mn@doi [\araa]
  {10.1146/annurev-astro-081817-051839}, \href
  {http://adsabs.harvard.edu/abs/2018ARA%26A..56...83B} {56, 83}

\bibitem[\protect\citeauthoryear{{Bastian}, {Lamers}, {de Mink}, {Longmore},
  {Goodwin}  \& {Gieles}}{{Bastian} et~al.}{2013}]{bastian13}
{Bastian} N.,  {Lamers} H.~J.~G.~L.~M.,  {de Mink} S.~E.,  {Longmore} S.~N.,
  {Goodwin} S.~P.,   {Gieles} M.,  2013, \mn@doi [\mnras]
  {10.1093/mnras/stt1745}, \href
  {https://ui.adsabs.harvard.edu/abs/2013MNRAS.436.2398B} {436, 2398}

\bibitem[\protect\citeauthoryear{{Bastian}, {Cabrera-Ziri}  \&
  {Salaris}}{{Bastian} et~al.}{2015}]{bastian15}
{Bastian} N.,  {Cabrera-Ziri} I.,   {Salaris} M.,  2015, \mn@doi [MNRAS]
  {10.1093/mnras/stv543}, \href
  {http://adsabs.harvard.edu/abs/2015MNRAS.449.3333B} {449, 3333}

\bibitem[\protect\citeauthoryear{{Bastian} et~al.,}{{Bastian}
  et~al.}{2019}]{bastian19}
{Bastian} N.,  et~al., 2019, \mn@doi [\mnras] {10.1093/mnrasl/slz130}, \href
  {https://ui.adsabs.harvard.edu/abs/2019MNRAS.489L..80B} {489, L80}

\bibitem[\protect\citeauthoryear{{Bastian} et~al.,}{{Bastian}
  et~al.}{2020}]{bastian20}
{Bastian} N.,  et~al., 2020, arXiv e-prints, \href
  {https://ui.adsabs.harvard.edu/abs/2020arXiv200303428B} {p. arXiv:2003.03428}

\bibitem[\protect\citeauthoryear{{Bellini}, {Anderson}  \& {Bedin}}{{Bellini}
  et~al.}{2011}]{bellini11}
{Bellini} A.,  {Anderson} J.,   {Bedin} L.~R.,  2011, \mn@doi [\pasp]
  {10.1086/659878}, \href
  {https://ui.adsabs.harvard.edu/abs/2011PASP..123..622B} {123, 622}

\bibitem[\protect\citeauthoryear{{Breen}}{{Breen}}{2018}]{breen18}
{Breen} P.~G.,  2018, \mn@doi [\mnras] {10.1093/mnrasl/sly169}, \href
  {http://adsabs.harvard.edu/abs/2018MNRAS.481L.110B} {481, L110}

\bibitem[\protect\citeauthoryear{{Cabrera-Ziri}, {Bastian}, {Davies}, {Magris},
  {Bruzual}  \& {Schweizer}}{{Cabrera-Ziri} et~al.}{2014}]{cabreraziri14}
{Cabrera-Ziri} I.,  {Bastian} N.,  {Davies} B.,  {Magris} G.,  {Bruzual} G.,
  {Schweizer} F.,  2014, \mn@doi [\mnras] {10.1093/mnras/stu764}, \href
  {https://ui.adsabs.harvard.edu/abs/2014MNRAS.441.2754C} {441, 2754}

\bibitem[\protect\citeauthoryear{{Cabrera-Ziri} et~al.,}{{Cabrera-Ziri}
  et~al.}{2015}]{cabrera_ziri15}
{Cabrera-Ziri} I.,  et~al., 2015, \mn@doi [\mnras] {10.1093/mnras/stv163},
  \href {https://ui.adsabs.harvard.edu/abs/2015MNRAS.448.2224C} {448, 2224}

\bibitem[\protect\citeauthoryear{{Cabrera-Ziri}, {Lardo}, {Davies}, {Bastian},
  {Beccari}, {Larsen}  \& {Hernandez}}{{Cabrera-Ziri}
  et~al.}{2016}]{cabreraziri16}
{Cabrera-Ziri} I.,  {Lardo} C.,  {Davies} B.,  {Bastian} N.,  {Beccari} G.,
  {Larsen} S.~S.,   {Hernandez} S.,  2016, \mn@doi [\mnras]
  {10.1093/mnras/stw1090}, \href
  {http://adsabs.harvard.edu/abs/2016MNRAS.460.1869C} {460, 1869}

\bibitem[\protect\citeauthoryear{{Cabrera-Ziri}, {Lardo}  \&
  {Mucciarelli}}{{Cabrera-Ziri} et~al.}{2019}]{cabreraziri19}
{Cabrera-Ziri} I.,  {Lardo} C.,   {Mucciarelli} A.,  2019, \mn@doi [\mnras]
  {10.1093/mnras/stz707}, \href
  {https://ui.adsabs.harvard.edu/abs/2019MNRAS.485.4128C} {485, 4128}

\bibitem[\protect\citeauthoryear{{Carretta} et~al.,}{{Carretta}
  et~al.}{2010}]{carretta10}
{Carretta} E.,  et~al., 2010, \mn@doi [\aap] {10.1051/0004-6361/201014924},
  \href {http://adsabs.harvard.edu/abs/2010A%26A...520A..95C} {520, A95}

\bibitem[\protect\citeauthoryear{{Casagrande} \& {VandenBerg}}{{Casagrande} \&
  {VandenBerg}}{2014}]{casagrande14}
{Casagrande} L.,  {VandenBerg} D.~A.,  2014, \mn@doi [\mnras]
  {10.1093/mnras/stu1476}, \href
  {https://ui.adsabs.harvard.edu/abs/2014MNRAS.444..392C} {444, 392}

\bibitem[\protect\citeauthoryear{{Chantereau}, {Salaris}, {Bastian}  \&
  {Martocchia}}{{Chantereau} et~al.}{2019}]{chantereau19}
{Chantereau} W.,  {Salaris} M.,  {Bastian} N.,   {Martocchia} S.,  2019, arXiv
  e-prints, \href {http://adsabs.harvard.edu/abs/2019arXiv190201806C} {}

\bibitem[\protect\citeauthoryear{{Choi}, {Dotter}, {Conroy}, {Cantiello},
  {Paxton}  \& {Johnson}}{{Choi} et~al.}{2016}]{2016ApJ...823..102C}
{Choi} J.,  {Dotter} A.,  {Conroy} C.,  {Cantiello} M.,  {Paxton} B.,
  {Johnson} B.~D.,  2016, \mn@doi [\apj] {10.3847/0004-637X/823/2/102}, \href
  {https://ui.adsabs.harvard.edu/abs/2016ApJ...823..102C} {823, 102}

\bibitem[\protect\citeauthoryear{{Colucci}, {Bernstein}  \& {Cohen}}{{Colucci}
  et~al.}{2014}]{colucci14}
{Colucci} J.~E.,  {Bernstein} R.~A.,   {Cohen} J.~G.,  2014, \mn@doi [\apj]
  {10.1088/0004-637X/797/2/116}, \href
  {https://ui.adsabs.harvard.edu/abs/2014ApJ...797..116C} {797, 116}

\bibitem[\protect\citeauthoryear{{D'Ercole}, {Vesperini}, {D'Antona},
  {McMillan}  \& {Recchi}}{{D'Ercole} et~al.}{2008}]{dercole08}
{D'Ercole} A.,  {Vesperini} E.,  {D'Antona} F.,  {McMillan} S.,   {Recchi} S.,
  2008, \mn@doi [MNRAS] {10.1111/j.1365-2966.2008.13915.x}, \href
  {http://adsabs.harvard.edu/abs/2008MNRAS.391..825D} {391, 825}

\bibitem[\protect\citeauthoryear{{Da Costa} \& {Hatzidimitriou}}{{Da Costa} \&
  {Hatzidimitriou}}{1998}]{dacosta98}
{Da Costa} G.~S.,  {Hatzidimitriou} D.,  1998, \mn@doi [\aj] {10.1086/300340},
  \href {https://ui.adsabs.harvard.edu/abs/1998AJ....115.1934D} {115, 1934}

\bibitem[\protect\citeauthoryear{{Dalessandro}, {Lapenna}, {Mucciarelli},
  {Origlia}, {Ferraro}  \& {Lanzoni}}{{Dalessandro}
  et~al.}{2016}]{dalessandro16}
{Dalessandro} E.,  {Lapenna} E.,  {Mucciarelli} A.,  {Origlia} L.,  {Ferraro}
  F.~R.,   {Lanzoni} B.,  2016, \mn@doi [ApJ] {10.3847/0004-637X/829/2/77},
  \href {http://adsabs.harvard.edu/abs/2016ApJ...829...77D} {829, 77}

\bibitem[\protect\citeauthoryear{{Decressin}, {Meynet}, {Charbonnel},
  {Prantzos}  \& {Ekstr{\"o}m}}{{Decressin} et~al.}{2007}]{decressin07}
{Decressin} T.,  {Meynet} G.,  {Charbonnel} C.,  {Prantzos} N.,   {Ekstr{\"o}m}
  S.,  2007, \mn@doi [Astron. Astrophys.] {10.1051/0004-6361:20066013}, \href
  {http://adsabs.harvard.edu/abs/2007A%26A...464.1029D} {464, 1029}

\bibitem[\protect\citeauthoryear{{Dotter}}{{Dotter}}{2016}]{2016ApJS..222....8D}
{Dotter} A.,  2016, \mn@doi [\apjs] {10.3847/0067-0049/222/1/8}, \href
  {https://ui.adsabs.harvard.edu/abs/2016ApJS..222....8D} {222, 8}

\bibitem[\protect\citeauthoryear{{Forbes} et~al.,}{{Forbes}
  et~al.}{2018}]{forbes18}
{Forbes} D.~A.,  et~al., 2018, \mn@doi [Proceedings of the Royal Society of
  London Series A] {10.1098/rspa.2017.0616}, \href
  {https://ui.adsabs.harvard.edu/abs/2018RSPSA.47470616F} {474, 20170616}

\bibitem[\protect\citeauthoryear{{Gaia Collaboration} et~al.,}{{Gaia
  Collaboration} et~al.}{2016}]{gaia2016}
{Gaia Collaboration} et~al., 2016, \mn@doi [\aap]
  {10.1051/0004-6361/201629512}, \href
  {https://ui.adsabs.harvard.edu/abs/2016A&A...595A...2G} {595, A2}

\bibitem[\protect\citeauthoryear{{Gaia Collaboration} et~al.,}{{Gaia
  Collaboration} et~al.}{2018}]{gaia2018}
{Gaia Collaboration} et~al., 2018, \mn@doi [\aap]
  {10.1051/0004-6361/201833051}, \href
  {https://ui.adsabs.harvard.edu/abs/2018A&A...616A...1G} {616, A1}

\bibitem[\protect\citeauthoryear{{Gieles} et~al.,}{{Gieles}
  et~al.}{2018}]{gieles18}
{Gieles} M.,  et~al., 2018, \mn@doi [\mnras] {10.1093/mnras/sty1059}, \href
  {http://adsabs.harvard.edu/abs/2018MNRAS.478.2461G} {478, 2461}

\bibitem[\protect\citeauthoryear{{Gilligan} et~al.,}{{Gilligan}
  et~al.}{2019}]{gilligan19}
{Gilligan} C.~K.,  et~al., 2019, arXiv e-prints, \href
  {http://adsabs.harvard.edu/abs/2019arXiv190401434G} {}

\bibitem[\protect\citeauthoryear{{Glatt} et~al.,}{{Glatt}
  et~al.}{2008}]{glatt08}
{Glatt} K.,  et~al., 2008, \mn@doi [AJ] {10.1088/0004-6256/136/4/1703}, \href
  {http://adsabs.harvard.edu/abs/2008AJ....136.1703G} {136, 1703}

\bibitem[\protect\citeauthoryear{{Gratton}, {Carretta}  \&
  {Bragaglia}}{{Gratton} et~al.}{2012}]{gratton12}
{Gratton} R.,  {Carretta} E.,   {Bragaglia} A.,  2012, \mn@doi [Astron.
  Astrophys.Rv] {10.1007/s00159-012-0050-3}, \href
  {http://adsabs.harvard.edu/abs/2012A\%26ARv..20...50G} {20, 50}

\bibitem[\protect\citeauthoryear{{Hollyhead} et~al.,}{{Hollyhead}
  et~al.}{2017}]{hollyhead17}
{Hollyhead} K.,  et~al., 2017, \mn@doi [MNRAS] {10.1093/mnrasl/slw179}, \href
  {http://adsabs.harvard.edu/abs/2017MNRAS.465L..39H} {465, L39}

\bibitem[\protect\citeauthoryear{{Hollyhead} et~al.,}{{Hollyhead}
  et~al.}{2018}]{hollyhead18}
{Hollyhead} K.,  et~al., 2018, \mn@doi [\mnras] {10.1093/mnras/sty230}, \href
  {http://adsabs.harvard.edu/abs/2018MNRAS.476..114H} {476, 114}

\bibitem[\protect\citeauthoryear{{Hollyhead} et~al.,}{{Hollyhead}
  et~al.}{2019}]{hollyhead19}
{Hollyhead} K.,  et~al., 2019, \mn@doi [\mnras] {10.1093/mnras/stz317}, \href
  {http://adsabs.harvard.edu/abs/2019MNRAS.484.4718H} {484, 4718}

\bibitem[\protect\citeauthoryear{{Husser}, {Wende-von Berg}, {Dreizler},
  {Homeier}, {Reiners}, {Barman}  \& {Hauschildt}}{{Husser}
  et~al.}{2013}]{husser13}
{Husser} T.~O.,  {Wende-von Berg} S.,  {Dreizler} S.,  {Homeier} D.,  {Reiners}
  A.,  {Barman} T.,   {Hauschildt} P.~H.,  2013, \mn@doi [\aap]
  {10.1051/0004-6361/201219058}, \href
  {https://ui.adsabs.harvard.edu/abs/2013A&A...553A...6H} {553, A6}

\bibitem[\protect\citeauthoryear{{Husser} et~al.,}{{Husser}
  et~al.}{2016}]{husser16}
{Husser} T.-O.,  et~al., 2016, \mn@doi [\aap] {10.1051/0004-6361/201526949},
  \href {https://ui.adsabs.harvard.edu/abs/2016A&A...588A.148H} {588, A148}

\bibitem[\protect\citeauthoryear{{Kamann}, {Wisotzki}  \& {Roth}}{{Kamann}
  et~al.}{2013}]{kamann13}
{Kamann} S.,  {Wisotzki} L.,   {Roth} M.~M.,  2013, \mn@doi [\aap]
  {10.1051/0004-6361/201220476}, \href
  {https://ui.adsabs.harvard.edu/abs/2013A&A...549A..71K} {549, A71}

\bibitem[\protect\citeauthoryear{{Kamann} et~al.,}{{Kamann}
  et~al.}{2018}]{kamann18}
{Kamann} S.,  et~al., 2018, \mn@doi [\mnras] {10.1093/mnras/sty1958}, \href
  {http://adsabs.harvard.edu/abs/2018MNRAS.480.1689K} {480, 1689}

\bibitem[\protect\citeauthoryear{{Kruijssen}}{{Kruijssen}}{2015}]{kruijssen15}
{Kruijssen} J.~M.~D.,  2015, \mn@doi [\mnras] {10.1093/mnras/stv2026}, \href
  {https://ui.adsabs.harvard.edu/abs/2015MNRAS.454.1658K} {454, 1658}

\bibitem[\protect\citeauthoryear{{Kurucz}}{{Kurucz}}{1970}]{1970SAOSR.309.....K}
{Kurucz} R.~L.,  1970, SAO Special Report, \href
  {https://ui.adsabs.harvard.edu/abs/1970SAOSR.309.....K} {309}

\bibitem[\protect\citeauthoryear{{Kurucz}}{{Kurucz}}{2005}]{2005MSAIS...8...14K}
{Kurucz} R.~L.,  2005, Memorie della Societa Astronomica Italiana Supplementi,
  \href {https://ui.adsabs.harvard.edu/abs/2005MSAIS...8...14K} {8, 14}

\bibitem[\protect\citeauthoryear{{Kurucz} \& {Avrett}}{{Kurucz} \&
  {Avrett}}{1981}]{1981SAOSR.391.....K}
{Kurucz} R.~L.,  {Avrett} E.~H.,  1981, SAO Special Report, \href
  {https://ui.adsabs.harvard.edu/abs/1981SAOSR.391.....K} {391}

\bibitem[\protect\citeauthoryear{{Kurucz} \& {Furenlid}}{{Kurucz} \&
  {Furenlid}}{1979}]{1979SAOSR.387.....K}
{Kurucz} R.~L.,  {Furenlid} I.,  1979, SAO Special Report, \href
  {https://ui.adsabs.harvard.edu/abs/1979SAOSR.387.....K} {387}

\bibitem[\protect\citeauthoryear{{Lagioia}, {Milone}, {Marino}  \&
  {Dotter}}{{Lagioia} et~al.}{2019}]{lagioia19}
{Lagioia} E.~P.,  {Milone} A.~P.,  {Marino} A.~F.,   {Dotter} A.,  2019,
  \mn@doi [\apj] {10.3847/1538-4357/aaf729}, \href
  {https://ui.adsabs.harvard.edu/abs/2019ApJ...871..140L} {871, 140}

\bibitem[\protect\citeauthoryear{{Lardo}, {Cabrera-Ziri}, {Davies}  \&
  {Bastian}}{{Lardo} et~al.}{2017}]{lardo17}
{Lardo} C.,  {Cabrera-Ziri} I.,  {Davies} B.,   {Bastian} N.,  2017, \mn@doi
  [\mnras] {10.1093/mnras/stx628}, \href
  {http://adsabs.harvard.edu/abs/2017MNRAS.468.2482L} {468, 2482}

\bibitem[\protect\citeauthoryear{{Lardo}, {Salaris}, {Bastian}, {Mucciarelli},
  {Dalessandro}  \& {Cabrera-Ziri}}{{Lardo} et~al.}{2018}]{lardo18}
{Lardo} C.,  {Salaris} M.,  {Bastian} N.,  {Mucciarelli} A.,  {Dalessandro} E.,
    {Cabrera-Ziri} I.,  2018, \mn@doi [\aap] {10.1051/0004-6361/201832999},
  \href {https://ui.adsabs.harvard.edu/abs/2018A&A...616A.168L} {616, A168}

\bibitem[\protect\citeauthoryear{{Larsen}, {Strader}  \& {Brodie}}{{Larsen}
  et~al.}{2012}]{larsen12}
{Larsen} S.~S.,  {Strader} J.,   {Brodie} J.~P.,  2012, \mn@doi [\aap]
  {10.1051/0004-6361/201219897}, \href
  {https://ui.adsabs.harvard.edu/abs/2012A&A...544L..14L} {544, L14}

\bibitem[\protect\citeauthoryear{{Larsen}, {Brodie}, {Forbes}  \&
  {Strader}}{{Larsen} et~al.}{2014a}]{larsen_et_al_14}
{Larsen} S.~S.,  {Brodie} J.~P.,  {Forbes} D.~A.,   {Strader} J.,  2014a,
  \mn@doi [\aap] {10.1051/0004-6361/201322672}, \href
  {https://ui.adsabs.harvard.edu/abs/2014A&A...565A..98L} {565, A98}

\bibitem[\protect\citeauthoryear{{Larsen}, {Brodie}, {Grundahl}  \&
  {Strader}}{{Larsen} et~al.}{2014b}]{larsen14}
{Larsen} S.~S.,  {Brodie} J.~P.,  {Grundahl} F.,   {Strader} J.,  2014b,
  \mn@doi [ApJ] {10.1088/0004-637X/797/1/15}, \href
  {http://adsabs.harvard.edu/abs/2014ApJ...797...15L} {797, 15}

\bibitem[\protect\citeauthoryear{{Latour} et~al.,}{{Latour}
  et~al.}{2019}]{latour19}
{Latour} M.,  et~al., 2019, \mn@doi [\aap] {10.1051/0004-6361/201936242}, \href
  {https://ui.adsabs.harvard.edu/abs/2019A&A...631A..14L} {631, A14}

\bibitem[\protect\citeauthoryear{{Li} \& {de Grijs}}{{Li} \& {de
  Grijs}}{2019}]{li_degrijs19}
{Li} C.,  {de Grijs} R.,  2019, \mn@doi [\apj] {10.3847/1538-4357/ab153b},
  \href {https://ui.adsabs.harvard.edu/abs/2019ApJ...876...94L} {876, 94}

\bibitem[\protect\citeauthoryear{{Marino} et~al.,}{{Marino}
  et~al.}{2019}]{marino19}
{Marino} A.~F.,  et~al., 2019, \mn@doi [\mnras] {10.1093/mnras/stz1415}, \href
  {https://ui.adsabs.harvard.edu/abs/2019MNRAS.487.3815M} {487, 3815}

\bibitem[\protect\citeauthoryear{{Martell}, {Smolinski}, {Beers}  \&
  {Grebel}}{{Martell} et~al.}{2011}]{martell11}
{Martell} S.~L.,  {Smolinski} J.~P.,  {Beers} T.~C.,   {Grebel} E.~K.,  2011,
  \mn@doi [\aap] {10.1051/0004-6361/201117644}, \href
  {https://ui.adsabs.harvard.edu/abs/2011A&A...534A.136M} {534, A136}

\bibitem[\protect\citeauthoryear{{Martocchia} et~al.,}{{Martocchia}
  et~al.}{2017}]{martocchia17}
{Martocchia} S.,  et~al., 2017, \mn@doi [MNRAS] {10.1093/mnras/stx660}, \href
  {http://adsabs.harvard.edu/abs/2017MNRAS.468.3150M} {468, 3150}

\bibitem[\protect\citeauthoryear{{Martocchia} et~al.,}{{Martocchia}
  et~al.}{2018a}]{martocchia18a}
{Martocchia} S.,  et~al., 2018a, \mn@doi [\mnras] {10.1093/mnras/stx2556},
  \href {https://ui.adsabs.harvard.edu/abs/2018MNRAS.473.2688M} {473, 2688}

\bibitem[\protect\citeauthoryear{{Martocchia} et~al.,}{{Martocchia}
  et~al.}{2018b}]{martocchia18b}
{Martocchia} S.,  et~al., 2018b, \mn@doi [\mnras] {10.1093/mnras/sty916}, \href
  {https://ui.adsabs.harvard.edu/abs/2018MNRAS.477.4696M} {477, 4696}

\bibitem[\protect\citeauthoryear{{Martocchia} et~al.,}{{Martocchia}
  et~al.}{2019}]{martocchia19}
{Martocchia} S.,  et~al., 2019, \mn@doi [\mnras] {10.1093/mnras/stz1596}, \href
  {https://ui.adsabs.harvard.edu/abs/2019MNRAS.487.5324M} {487, 5324}

\bibitem[\protect\citeauthoryear{{McLaughlin} \& {van der Marel}}{{McLaughlin}
  \& {van der Marel}}{2005}]{mclaughlin05}
{McLaughlin} D.~E.,  {van der Marel} R.~P.,  2005, \mn@doi [\apjs]
  {10.1086/497429}, \href {http://adsabs.harvard.edu/abs/2005ApJS..161..304M}
  {161, 304}

\bibitem[\protect\citeauthoryear{{Meurer} et~al.,}{{Meurer}
  et~al.}{2003}]{meurer03}
{Meurer} G.~R.,  et~al., 2003, in {Blades} J.~C.,  {Siegmund} O. H.~W.,  eds,
  Society of Photo-Optical Instrumentation Engineers (SPIE) Conference Series
  Vol. 4854, \procspie. pp 507--514, \mn@doi{10.1117/12.460259}

\bibitem[\protect\citeauthoryear{{Milone} et~al.,}{{Milone}
  et~al.}{2012}]{milone12}
{Milone} A.~P.,  et~al., 2012, \mn@doi [\aap] {10.1051/0004-6361/201016384},
  \href {http://adsabs.harvard.edu/abs/2012A%26A...540A..16M} {540, A16}

\bibitem[\protect\citeauthoryear{{Milone} et~al.,}{{Milone}
  et~al.}{2015}]{milone15b}
{Milone} A.~P.,  et~al., 2015, \mn@doi [\apj] {10.1088/0004-637X/808/1/51},
  \href {https://ui.adsabs.harvard.edu/abs/2015ApJ...808...51M} {808, 51}

\bibitem[\protect\citeauthoryear{{Milone} et~al.,}{{Milone}
  et~al.}{2017}]{milone17}
{Milone} A.~P.,  et~al., 2017, \mn@doi [\mnras] {10.1093/mnras/stw2531}, \href
  {https://ui.adsabs.harvard.edu/abs/2017MNRAS.464.3636M} {464, 3636}

\bibitem[\protect\citeauthoryear{{Moffat}}{{Moffat}}{1969}]{moffat69}
{Moffat} A.~F.~J.,  1969, \aap, \href
  {https://ui.adsabs.harvard.edu/abs/1969A&A.....3..455M} {3, 455}

\bibitem[\protect\citeauthoryear{{Montegriffo}, {Ferraro}, {Fusi Pecci}  \&
  {Origlia}}{{Montegriffo} et~al.}{1995}]{montegriffo1995}
{Montegriffo} P.,  {Ferraro} F.~R.,  {Fusi Pecci} F.,   {Origlia} L.,  1995,
  \mn@doi [\mnras] {10.1093/mnras/276.3.739}, \href
  {https://ui.adsabs.harvard.edu/abs/1995MNRAS.276..739M} {276, 739}

\bibitem[\protect\citeauthoryear{{Mucciarelli}, {Carretta}, {Origlia}  \&
  {Ferraro}}{{Mucciarelli} et~al.}{2008}]{mucciarelli08}
{Mucciarelli} A.,  {Carretta} E.,  {Origlia} L.,   {Ferraro} F.~R.,  2008,
  \mn@doi [AJ] {10.1088/0004-6256/136/1/375}, \href
  {http://adsabs.harvard.edu/abs/2008AJ....136..375M} {136, 375}

\bibitem[\protect\citeauthoryear{{Mucciarelli}, {Origlia}, {Ferraro}  \&
  {Pancino}}{{Mucciarelli} et~al.}{2009}]{mucciarelli09}
{Mucciarelli} A.,  {Origlia} L.,  {Ferraro} F.~R.,   {Pancino} E.,  2009,
  \mn@doi [ApJ] {10.1088/0004-637X/695/2/L134}, \href
  {http://adsabs.harvard.edu/abs/2009ApJ...695L.134M} {695, L134}

\bibitem[\protect\citeauthoryear{{Mucciarelli}, {Dalessandro}, {Ferraro},
  {Origlia}  \& {Lanzoni}}{{Mucciarelli} et~al.}{2014}]{mucciarelli14}
{Mucciarelli} A.,  {Dalessandro} E.,  {Ferraro} F.~R.,  {Origlia} L.,
  {Lanzoni} B.,  2014, \mn@doi [\apjl] {10.1088/2041-8205/793/1/L6}, \href
  {http://adsabs.harvard.edu/abs/2014ApJ...793L...6M} {793, L6}

\bibitem[\protect\citeauthoryear{{Neumayer}, {Seth}  \& {Boeker}}{{Neumayer}
  et~al.}{2020}]{neumayer20}
{Neumayer} N.,  {Seth} A.,   {Boeker} T.,  2020, arXiv e-prints, \href
  {https://ui.adsabs.harvard.edu/abs/2020arXiv200103626N} {p. arXiv:2001.03626}

\bibitem[\protect\citeauthoryear{{Niederhofer} et~al.,}{{Niederhofer}
  et~al.}{2017a}]{niederhofer17a}
{Niederhofer} F.,  et~al., 2017a, \mn@doi [MNRAS] {10.1093/mnras/stw2269},
  \href {http://adsabs.harvard.edu/abs/2017MNRAS.464...94N} {464, 94}

\bibitem[\protect\citeauthoryear{{Niederhofer} et~al.,}{{Niederhofer}
  et~al.}{2017b}]{niederhofer17b}
{Niederhofer} F.,  et~al., 2017b, \mn@doi [MNRAS] {10.1093/mnras/stw3084},
  \href {http://adsabs.harvard.edu/abs/2017MNRAS.465.4159N} {465, 4159}

\bibitem[\protect\citeauthoryear{{Paxton}, {Bildsten}, {Dotter}, {Herwig},
  {Lesaffre}  \& {Timmes}}{{Paxton} et~al.}{2011}]{2011ApJS..192....3P}
{Paxton} B.,  {Bildsten} L.,  {Dotter} A.,  {Herwig} F.,  {Lesaffre} P.,
  {Timmes} F.,  2011, \mn@doi [\apjs] {10.1088/0067-0049/192/1/3}, \href
  {https://ui.adsabs.harvard.edu/abs/2011ApJS..192....3P} {192, 3}

\bibitem[\protect\citeauthoryear{{Prantzos} \& {Charbonnel}}{{Prantzos} \&
  {Charbonnel}}{2006}]{prantzos06}
{Prantzos} N.,  {Charbonnel} C.,  2006, \mn@doi [\aap]
  {10.1051/0004-6361:20065374}, \href
  {https://ui.adsabs.harvard.edu/abs/2006A&A...458..135P} {458, 135}

\bibitem[\protect\citeauthoryear{{Sakari} et~al.,}{{Sakari}
  et~al.}{2016}]{sakari16}
{Sakari} C.~M.,  et~al., 2016, \mn@doi [\apj] {10.3847/0004-637X/829/2/116},
  \href {https://ui.adsabs.harvard.edu/abs/2016ApJ...829..116S} {829, 116}

\bibitem[\protect\citeauthoryear{{Salaris} et~al.,}{{Salaris}
  et~al.}{2020}]{salaris20}
{Salaris} M.,  et~al., 2020, \mn@doi [\mnras] {10.1093/mnras/staa089}, \href
  {https://ui.adsabs.harvard.edu/abs/2020MNRAS.492.3459S} {492, 3459}

\bibitem[\protect\citeauthoryear{{Saracino} et~al.,}{{Saracino}
  et~al.}{2019}]{saracino19}
{Saracino} S.,  et~al., 2019, \mn@doi [\mnras] {10.1093/mnrasl/slz135}, \href
  {https://ui.adsabs.harvard.edu/abs/2019MNRAS.489L..97S} {489, L97}

\bibitem[\protect\citeauthoryear{{Saracino} et~al.,}{{Saracino}
  et~al.}{2020a}]{saracino20}
{Saracino} S.,  et~al., 2020a, arXiv e-prints, \href
  {https://ui.adsabs.harvard.edu/abs/2020arXiv200301780S} {p. arXiv:2003.01780}

\bibitem[\protect\citeauthoryear{{Saracino} et~al.,}{{Saracino}
  et~al.}{2020b}]{saracino20b}
{Saracino} S.,  et~al., 2020b, arXiv e-prints, \href
  {https://ui.adsabs.harvard.edu/abs/2020arXiv200903320S} {p. arXiv:2009.03320}

\bibitem[\protect\citeauthoryear{{Schiavon}, {Caldwell}, {Conroy}, {Graves},
  {Strader}, {MacArthur}, {Courteau}  \& {Harding}}{{Schiavon}
  et~al.}{2013}]{schiavon13}
{Schiavon} R.~P.,  {Caldwell} N.,  {Conroy} C.,  {Graves} G.~J.,  {Strader} J.,
   {MacArthur} L.~A.,  {Courteau} S.,   {Harding} P.,  2013, \mn@doi [\apjl]
  {10.1088/2041-8205/776/1/L7}, \href
  {http://adsabs.harvard.edu/abs/2013ApJ...776L...7S} {776, L7}

\bibitem[\protect\citeauthoryear{{Stetson}}{{Stetson}}{1987}]{stetson87}
{Stetson} P.~B.,  1987, \mn@doi [PASP] {10.1086/131977}, \href
  {http://adsabs.harvard.edu/abs/1987PASP...99..191S} {99, 191}

\bibitem[\protect\citeauthoryear{{Stetson}}{{Stetson}}{1994}]{stetson94}
{Stetson} P.~B.,  1994, \mn@doi [PASP] {10.1086/133378}, \href
  {http://adsabs.harvard.edu/abs/1994PASP..106..250S} {106, 250}

\bibitem[\protect\citeauthoryear{{Weilbacher} et~al.,}{{Weilbacher}
  et~al.}{2020}]{weilbacher2020}
{Weilbacher} P.~M.,  et~al., 2020, arXiv e-prints, \href
  {https://ui.adsabs.harvard.edu/abs/2020arXiv200608638W} {p. arXiv:2006.08638}

\bibitem[\protect\citeauthoryear{{Zhang}, {de Grijs}, {Li}  \& {Wu}}{{Zhang}
  et~al.}{2018}]{zhang18}
{Zhang} H.,  {de Grijs} R.,  {Li} C.,   {Wu} X.,  2018, \mn@doi [\apj]
  {10.3847/1538-4357/aaa428}, \href
  {https://ui.adsabs.harvard.edu/abs/2018ApJ...853..186Z} {853, 186}

\bibitem[\protect\citeauthoryear{{de Mink}, {Pols}, {Langer}  \& {Izzard}}{{de
  Mink} et~al.}{2009}]{demink09}
{de Mink} S.,  {Pols} O.,  {Langer} N.,   {Izzard} R.,  2009, \mn@doi [Astron.
  Astrophys.] {10.1051/0004-6361/200913205}, \href
  {http://adsabs.harvard.edu/abs/2009A%26A...507L...1D} {507, L1}

\makeatother
\end{thebibliography}






\bsp	
\label{lastpage}

%

\end{document}